\documentclass[journal,onecolumn,10pt,twoside]{IEEEtranTCOM}
\usepackage{amssymb}
\usepackage{pifont}
\usepackage{bbding}

\usepackage[cmex10]{amsmath}
\usepackage{array, setspace}
\usepackage{mdwmath}
\usepackage{mdwtab}
\usepackage{booktabs}
\usepackage[usenames,dvipsnames]{color}
\usepackage{cases}
\usepackage{multirow, lscape}
\usepackage[ampersand]{easylist}
\usepackage{graphicx}
\usepackage{color}
\usepackage{cite}
\usepackage{graphicx, subfigure}
\usepackage[cmex10]{amsmath}
\usepackage{esint}
\usepackage{amssymb}
\usepackage{cases}
\usepackage{algorithm}
\usepackage{array}
\usepackage{epstopdf}
\usepackage{multirow}
\usepackage{caption}
\usepackage{amsthm}
\usepackage{algpseudocode}
\usepackage{threeparttable}
 \theoremstyle{remark}

\theoremstyle{}
\newtheorem{theorem}{Theorem}
\theoremstyle{}

\newtheorem{lemma}{Lemma}
\theoremstyle{}

\theoremstyle{remark}
\newtheorem{example}{Example}

\normalsize

\ifCLASSINFOpdf
\else
\fi
\begin{document}
\begin{spacing}{1.0}

\title{On the Coded Caching Delivery Design over Wireless Networks}

\author{Lei Zheng, Qifa~Yan, Qingchun~Chen, Xiaohu~Tang}

\maketitle
\begin{abstract}
Coded caching scheme is a promising technique to migrate the network burden in peak hours, which attains more prominent gains than the
uncoded caching. The coded caching scheme can be classified into two types, namely, the centralized
and the decentralized scheme, according to whether the placement procedures are carefully designed or
operated at random. However, most of the previous analysis assumes that the connected links between server and users are error-free. 
In this paper, we explore the coded caching based delivery design in wireless networks, where
all the connected wireless links are different. For both centralized and decentralized cases, we proposed two delivery schemes, namely, the \emph{orthogonal delivery} scheme and the \emph{concurrent delivery} scheme. We focus on the transmission time slots spent on
satisfying the system requests, and prove that
for both the centralized and the decentralized cases, the
\emph{concurrent delivery} always outperforms \emph{orthogonal delivery} scheme.
Furthermore, for the \emph{orthogonal delivery} scheme, we derive the gap in terms of transmission time between the decentralized and centralized
case, which is essentially no more than 1.5.

\end{abstract}
\IEEEpeerreviewmaketitle
\section{Introduction}

With the dramatic increasing demands for multimedia services, a huge burden is placed on
wireless multimedia transmission. Recently,
caching is considered to be one promising
technique to mitigate this burden, for it is increasingly cheap. Generally,
the popular contents are pre-fetched and
cached at the local storage of users, once the cached contents are requested by the
associated user, it can be delivered locally, thereby, the remote server only needs to transmit the
non-cached contents. With the help of caches, the amount of transmissions by
server is greatly reduced.

In their seminal work, Maddah-Ali and Niesen proved that the amount of transmission at server can be further
reduced by exploring coding into the multicast streaming according to the content caching at different users, which is referred to as coded
caching scheme\cite{Ali2014}\cite{Ali2015}.
The system consists of one single server with a file library, multiple users
each equipped with an identical cache, as well as the associated error-free and shared links. The system generally operates in two phases, i.e., placement phase and delivery phase. The placement phase occurs in the off peak hours (such as midnight) and operates without the information of  different users' requests. Each user's cache is filled up with a fraction of each file due to the cache size limit. The delivery phase operates in peak hours. Based on the user requests and the knowledge of the cached contents at users, the server responds by multicasting coded segments to a set of users.
The corresponding user is able to recover the requested file based on the received coded segments and the local cached contents.
By applying the coded caching scheme, the multicast opportunities can be created to improve the transmission efficiency significantly. Now multiple users can be served by sending one coded segment from server simultaneously, ever though the relative requests are distinct.

Basically, the coded caching scheme can be classified into centralized \cite{Ali2014} and the decentralized scheme \cite{Ali2015}, according to whether the placement phase is carefully designed or randomly handled. In the centralized scheme, the placement phase are coordinated by a central server among all the users to create more coded multicasting opportunities. While in the decentralized case, the placement phase are accomplished at random for each user. Now the coded multicasting opportunities depend on the cached contents and different users' requests. In most of the previous analysis, it is usually assumed that all users connect to the server through error-free shared links. Nonetheless, this assumption is no longer true in wireless networks. Elia et al. considered the wireless coded caching from a topological perspective \cite{Elia2016}, but only strong or weak abstract channels were considered.
The noisy broadcast network with caching was studied in \cite{Timo2016}, but the essential coded caching scheme was not yet taken into considerations.
The single shared link was extended into 2 and 3 parallel partially shared links in \cite{Caire2016}, the order-optimal rate and maximal delay region was investigated, while, the connected link was performed as error-free and shared.

In this paper, we explore the coded caching (both centralized and decentralized cases are
included) over wireless Gaussian Broadcast Channels (GBCs), and propose two delivery schemes,
namely: \emph{orthogonal delivery} scheme and \emph{concurrent delivery} scheme.
The transmission time slots defined as the time spent on satisfying all the users
requests is investigated.
Theoretically, both for centralized and decentralized placement cases, we prove that
\emph{concurrent delivery} scheme outperforms \emph{orthogonal delivery}. While, the
simulation results indicate that this advantage is gradually vanishing with the increasing cache size.
Specifically, when the same \emph{orthogonal delivery} scheme is assumed, the gap of transmission time slots between the decentralized scheme and the centralized scheme is turned out to be no more than 1.5,
which is in accordance with the gap of data rate over error-free shared channels in \cite{qifa2016}.
Our analysis in this paper reveals that the transmission time slots essentially depend on the traffic load even for wireless broadcast networks.

The remainder of this paper is organized as follows. In Section
 \uppercase\expandafter{\romannumeral2}, we introduce the system model and the
 preliminaries. Section \uppercase\expandafter{\romannumeral3} and Section
 \uppercase\expandafter{\romannumeral4} present the proposed \emph{orthogonal delivery}
 and \emph{concurrent delivery} schemes for the centralized and the decentralized case respectively.
 We conclude our work in Section \uppercase\expandafter{\romannumeral5}.
 \begin{figure}
  \centering
  \includegraphics[width=3.5in]{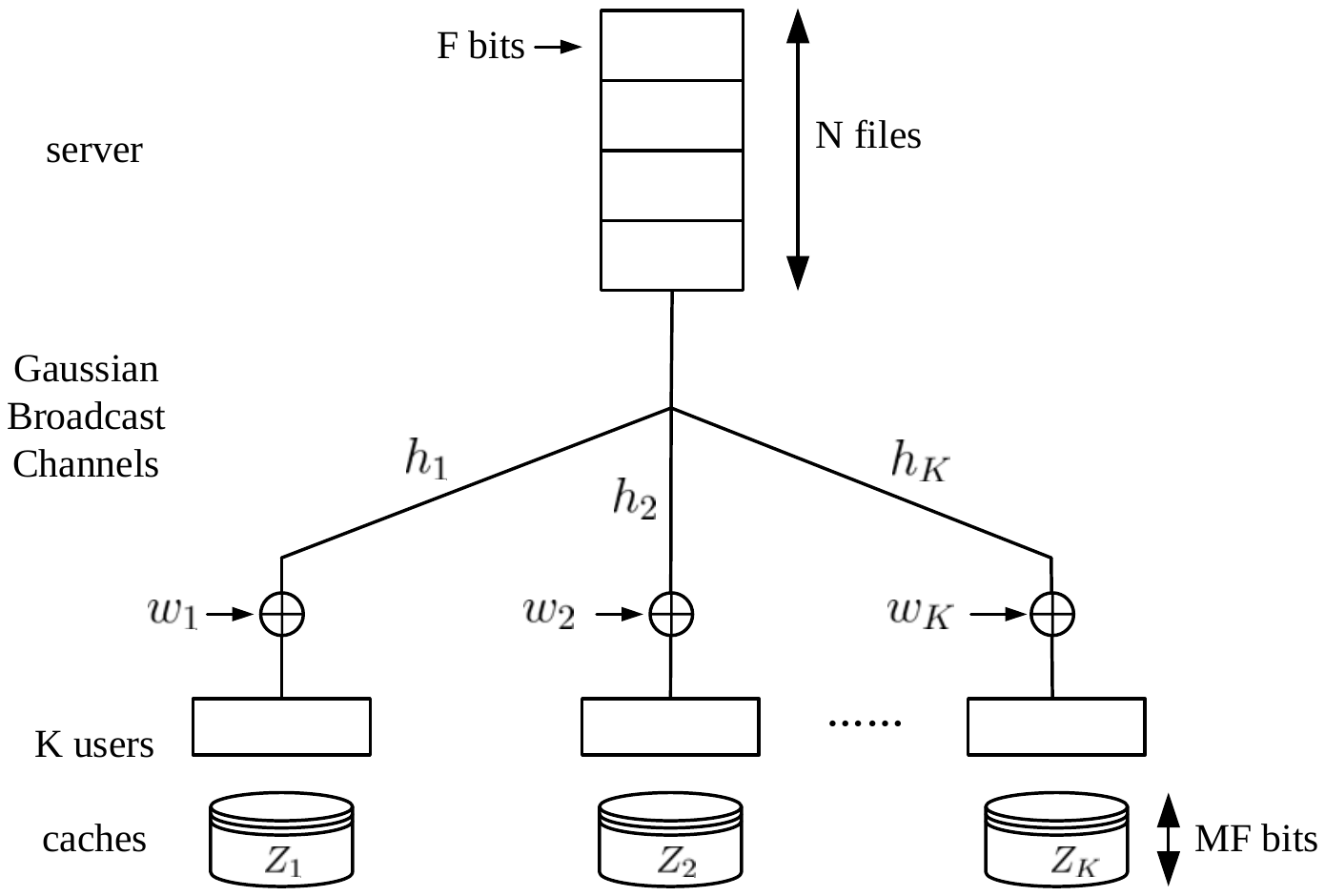}\\
  \caption{Coded caching in wireless network}\label{fig:system}
\end{figure}

\section{Network Model and Preliminaries}


\subsection{Network Model}
Consider a system with a single server caching $N$ files
$\mathcal{W}=\{W_1,W_2,\cdots,W_N\}$, each with the same length $F$ bits, and $K$
users $\mathcal{K}=\{U_1,U_2,\cdots,U_K\}$ are connected to the server through a wireless Gaussian broadcast channel. It is assumed that each user $k$ has an isolated cache of size $MF$ bits denoted by $Z_k$. As depicted in Fig. \ref{fig:system},
the channel for user $k$ is suffered from additive white gaussian noise
$w_k$ and has a distinct channel gain $h_k$,
both the noise $w_k$ and the channel gain $h_k$ are assumed
 to be complex Gaussian with zero mean and unit variance. Without loss in generality, assume that $0\leq |h_1|\leq |h_2|\leq \cdots\leq |k_K|<\infty$.
 For each user $k$, the received signal $y_k$ from the server is given by
 \begin{align}
 y_k=h_kx+w_k, \quad k=1,2,\cdots,K,
 \end{align}
 where the transmitted power by the server is assumed to be unity, namely, $\mathbf{E}[|x|^2]= 1$.

  The whole system operates in two phases, the placement phase followed by delivery phase.
In the placement phase, each user $k$ is available to access to the server and fill up its
cache $Z_k$ with some contents out of the file library without having the knowledge of users' requests,
 which occurs in off peak hours. On the contrary, the delivery phase takes place in the
 peak hours. During this phase, each user $k$ requests file $W_{d_k}$
 from $\mathcal{W}$. Generally speaking, different users' requests are assumed to be independently identically distributed (i.i.d.). Meanwhile, it is assumed that the channel gain $h_k$ can be estimated at the $k$th user and feedback to the server. After combining the users' requests $\mathcal{D}=\{d_1,d_2,\cdots,d_K\}$, the caching
  contents $\mathcal{Z}=\{Z_1,Z_2,\cdots,Z_K\}$ and channel state information
   $\mathcal{H}=\{h_1,h_2,\cdots,h_K\}$, the server will transmit coded signals of $R^{\left(\mathcal{H},M\right)}F$
   bits to the users. Each user $k$ is able to restore the requested file $W_{d_k}$ from the received coded signals and the pre-cached contents.
   Given $N,M,K$ and the channel state information $\mathcal{H}$, $R^{\left(\mathcal{H},M\right)}F$
   is said to be achievable if it is achievable for each request $\mathcal{D}$. In our following discussion,
    it is referred to as transmission load, notably, due to the identical file length $F$ bits,
all of our following considerations are normalized by $F$.

 The channel capacity of the aformentioned coded caching system over GBCs is denoted by $\mathcal{C}$. Let $\mathcal{T}$ indicate the transmission time slots
 spent on satisfying all the users' requests:
 \begin{align}
 \mathcal{T}\left(\mathcal{H},M\right)=\frac{R^{\left(\mathcal{H},M\right)}}{\mathcal{C}}.
 \end{align}

 For the given parameters $K,N,M,\mathcal{D}$ and $\mathcal{H}$, the transmission time slots
 should be expected as small as possible. Therefore, in this paper, we propose two delivery schemes to minimize $\mathcal{T}$,
 namely, \emph{orthogonal delivery }
 and \emph{concurrent delivery} scheme for both the centralized and decentralized schemes.
\subsection{Coded Caching based Delivery Scheme}

1) \emph{Orthogonal Delivery Scheme}

\begin{figure}
  \centering
  \includegraphics[width=3in]{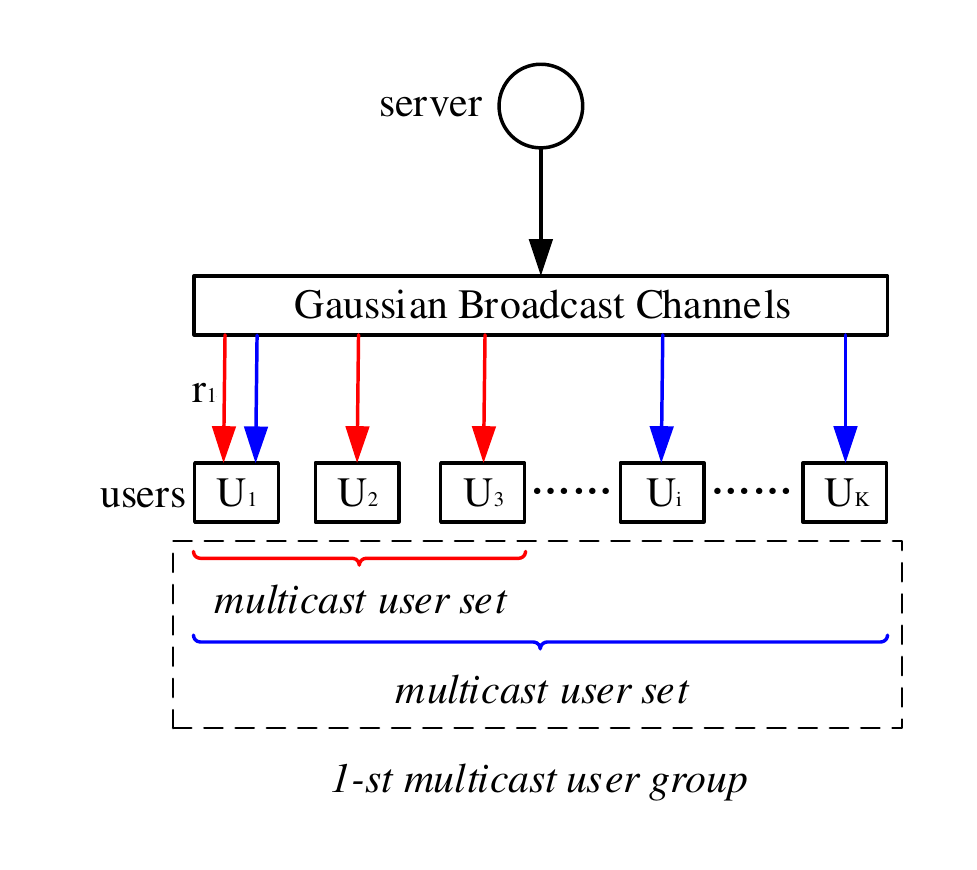}\\
  \caption{Orthogonal delivery illustration. red and blue arrows represent two distinct multicast user sets, while, they have the same worst user $U_1$. For $U_1$, the two multicast user sets may form the 1st multicast user group, and served with the maximum rate $r_1$ constrained by the worst user $U_1$'s channel capacity. }\label{fig:orthogonal_channel_model}
\end{figure}

In the coded caching strategy, multicasting opportunities can be achieved, which
implies that a set of users can be served by one transmission simultaneously. In our discussion, the
set of simultaneously serving users is referred to as the \emph{multicast user set}. As shown in Fig. \ref{fig:orthogonal_channel_model}, the users labeled with red and blue arrows form two multicast user sets. Apparently, the worst user for the two multicast user sets are the same $U_1$, hence, the maximum transmission rate for the two multicast user sets is constrained by the channel capacity $r_1$ of the worst user $U_1$.
 Moreover, the group of multicast user sets, which is dominated by the same worst user $k$ may constitute a \emph{kth multicast user group} and can be served with the same maximum transmission rate, e.g., the 1st multicast user group constrained by $U_1$ with the maximum transmission rate $r_1$.

For multicast and asymmetric channel state, the better users are able to decode the
signal transmitted to worse users, thereby, the worst users are always preferentially served. During the transmissions, the server sends coded segments to only one multicast user group per transmission, with the channel capacity of the worst user as the maximum transmission rate,
e.g., the 1st multicast user group is served with rate $r_1$ firstly,
then the 2nd multicast user group with rate $r_2$ follows, and so on. The system transmission time slots are the sum of each transmission.

2) \emph{Concurrent Delivery Scheme}

For the concurrent delivery scheme, all the multicast user groups are served simultaneously, however, the transmission rate is not the corresponding channel capacity any more, which
should be carefully allocated.

Similarly, the users may constitute different multicast user groups, to make full use of all the available channels, the multicast user groups tend to be served simultaneously, which is supported by the superposition code at the server and the Successive Interference Cancellation (SIC)
at the user side, for the optimal capacity achievability of the code and decode strategies. Traditionally, for the transmit signal is the linear superposition of the
signals to all the multicast user groups, firstly, the worse user treats the signal of
better user as noise and decodes its signal, secondly, the better user performs SIC,
it decodes the transmit signal corresponding to worse user, and then proceeds to subtract
the signal of worse user and decodes its own signal. Even though, the better user can decode any signals that worse users can successfully decode, they will be discarded for useless. Whereas, for concurrent delivery scheme, the overall decoded signals for each user are
valuable to decode the requested segments.

\begin{figure}
  \centering
  \includegraphics[width=3.5in]{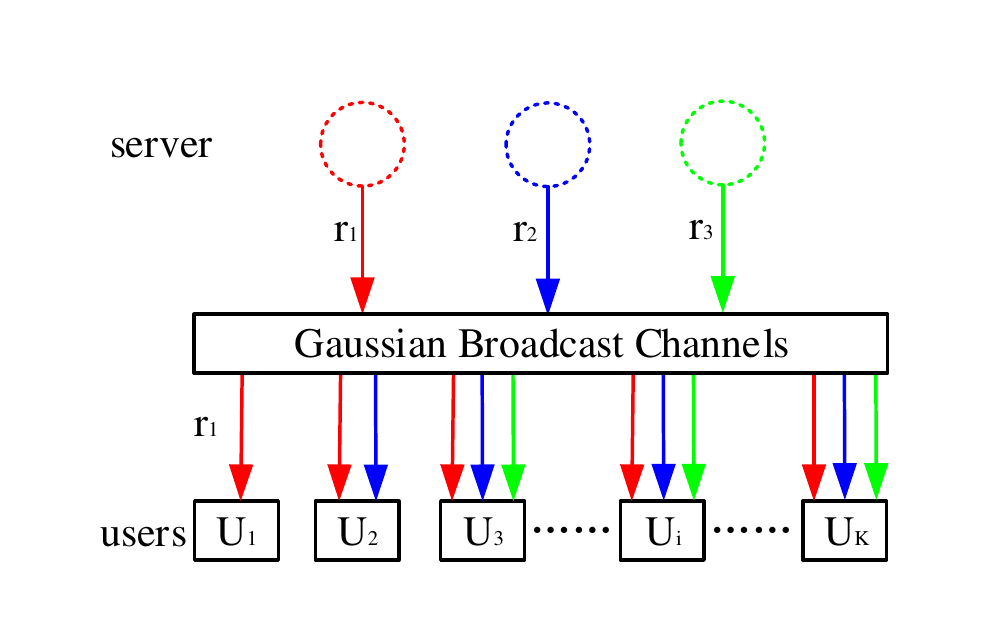}\\
  \caption{Concurrent delivery scheme. Red, blue and green arrows denote
  different transmit signals corresponding to the 1st, 2nd and 3rd multicast
  user groups. The corresponding transmission rate $r_1,r_2$ and $r_3$
  should satisfy the GBC capacity constraint.}\label{fig:concurrent_channel_model}
\end{figure}

The Fig. \ref{fig:concurrent_channel_model} depicts the
concurrent delivery procedures. The users labeled with different colored arrows formed into
different multicast user groups, the signal corresponding to 1st multicast user protruded by red can be decoded by each user, the blue colored signal can be decoded
by the following $K-1$ better users, and the green signal can be decoded by $K-2$ better users, and so on, wherein, the transmission rate for the three signals are $r_1,r_2,r_3$ respectively,
which should be carefully allocated subjected to the GBC capacity.

During the delivery phase, the signals transmit to all the multicast user groups, are
supercoded and send out together by the server, with respective transmission rate.
Interestingly, the single server can be regarded as multiple virtual servers, and the connected links to each multicast user group are extended into multiple \emph{parallel partially wireless links},
such that, each multicast user group corresponds to a partially wireless link.
Due to the adopted coded caching, each user is able to
extract the requested segments from its decoded signals. Consequently, the system
transmission time slots depend on the longest time spent on extracting quested segments of
each user.

\subsection{Preliminary}
The coded caching is applied over the GBCs, where channel capacity $\mathcal{C}$
is constrained by the GBC capacity region $\mathcal{C}_{BC}$. Consider the general case that the channels
are asymmetric, such as: $|h_1|\leq |h_2|\leq \cdots\leq |h_K|$, the GBC capacity region
$\mathcal{C}_{BC}^{\star}$ takes the form \cite{DavidTse}:
\begin{align}
\mathcal{C}_{BC}^{\star}=\left\{\mathbf{r}~\left|~r_i\leq\log_2\left(1+\frac{\alpha_i|h_i|^2}{1+(\sum \limits_{j>i}\alpha_j)|h_i|^2}\right),
0\leq\alpha_i\leq 1,\sum_{i=1}^K\alpha_i=1, r_i\geq 0,\forall i\right.\right\},
\end{align}
where $\alpha_i$ is the power allocation factor depends on the specific power allocation scheme,
meanwhile, the bandwidth, noise power and total transmit power are assumed to be unity 1.
The above capacity region also takes another equivalent form elaborated in Lemma \ref{Cbc}:
\begin{lemma}\label{Cbc}
The Gaussian broadcast channel capacity region takes the equivalent form as $\mathcal{C}_{BC}$:
\begin{align}
\mathcal{C}_{BC}=\left\{\mathbf{r}~\left|~\sum_{i=1}^K\left(\frac{1}{|h_i|^2}-\frac{1}{|h_{i+1}|^2}\right)
2^{\sum \limits_{j=1}^ir_j}-\frac{1}{|h_1|^2}\leq 1,
r_i\geq0,\forall ~i,\frac{1}{|h_{K+1}|^2}=0\right.\right\}.\notag
\end{align}
\end{lemma}
The proof of Lemma \ref{Cbc} is presented in Appendix A.
For characterizing the capacity region only with the unknown channel state, we take the equivalent form $\mathcal{C}_{BC}$
for description.

\section{Delivery Design for centralized Scheme}
For centralized case, the placement phase stays the same with Ali-Niesen's centralized placement procedures
\cite{Ali2014}. Each file is equally divided into $\binom{K}{t}$ segments and each has size $F/\binom{K}{t}$
bits, where $t=KM/N \in \mathbb{N}^+$ indicates the ratio
that all the users' caches to the total file size. Each user chooses $M/N$ segments of each file to store,
where the segments should be carefully designed, which constrained by the cache size.
This centralized placement phase guarantees that
each coded segment can be multicast to $(t+1)$ users, whereby, the set of users is termed \emph{multicast user set} in the
following. The Algorithm \ref{alg:centralizedplacement} describes this centralized placement procedures:
\begin{algorithm}
\caption{Centralized Placement}
\begin{algorithmic}[1]
\Procedure {Placement Procedures}{}
\State $t \leftarrow KM/N$
\State $\mathcal{I}\leftarrow \{\textit{T}\subset [K]:|\textit{T}|=t\}$
\For {$n\in [N]$}
\State split each file $W_n$ into $(W_{n,\textit{T}}:\textit{T}\subset \mathcal{I})$ of equal size
\EndFor
\For{$k\in [K]$}
\State $Z_k\leftarrow (W_{n,\textit{T}}:n\in [N],\textit{T}\subset \mathcal{I},k\in \textit{T})$
\EndFor
\EndProcedure
\end{algorithmic}\label{alg:centralizedplacement}
\end{algorithm}

In the following delivery phase, we propose two delivery scheme: \emph{orthogonal delivery} scheme and
\emph{concurrent delivery} scheme, to minimize the transmission time slots.

\subsection{orthogonal delivery scheme}

For given $K,M,N$, there are $\binom{K}{t+1}$ multicast user sets, in which the  \emph{$i$th
 multicast user group} denoted by $\mathcal{G}_i$ $\left(1\leq i\leq (K-t)\right)$, is constituted by $\binom{K-i}{t}$ multicast user sets.
To guarantee that each user successfully decode, the maximal
transmission rate to the \emph{$i$th multicast user group} is limited by $\log_2(1+|h_i|^2)$. And, each multicast user group is served one after another.

Based on the above description, we have the total transmission time slots of orthogonal delivery scheme for centralized case,
described as Theorem \ref{the:tco}:
\begin{theorem}\label{the:tco}
The transmission time slots of orthogonal delivery scheme for centralized case is denoted by $\mathcal{T}_{c,o}(\mathcal{H},M)$:
 \begin{align}
 \mathcal{T}_{c,o}(\mathcal{H},M)=\frac{1}{\binom{K}{t}}\sum \limits_{i=1}^{K-t}\frac{\binom{K-i}{t}}{\log_2(1+|h_i|^2)}.
 \end{align}
\end{theorem}

For completeness, the Algorithm \ref{alg:ortdelivery} describes the orthogonal delivery procedures based on the placement
procedures in Algorithm \ref{alg:centralizedplacement}.
\begin{algorithm}
\caption{Centralized Delivery}
\begin{algorithmic}[1]
\Procedure {Orthogonal Delivery procedures}{}
\State $t \leftarrow KM/N$
\For {$k=[K-t]$}
\State $\mathcal{S}\leftarrow \left\{S=\{s_1,\cdots,s_{t+1}\}\subset [K],s_1=k\right\}$
\State send $\mathbf{X}_{\mathcal{D}}\leftarrow (\oplus_{n\in S} W_{d_n,S\backslash\{n\}}:S\in \mathcal{S})$
with rate $\log_2(1+|h_k|^2)$
\EndFor
\EndProcedure
\end{algorithmic}\label{alg:ortdelivery}
\end{algorithm}

Where $W_{d_n,S\backslash\{n\}}$ presents the requested file segment by user $n$, which is missing at $n$,
while stored at the remaining users in $S$, and $\oplus$ denotes the bitwise operation.
To give the details of the orthogonal delivery procedures, we take the Example \ref{exa:ortde} for illustration.

\begin{example}\label{exa:ortde}
As shown in Fig. \ref{fig:centralizedo}, it is assumed the server holds $N=4$ files denoted by
$ \mathcal{W}=\{A,B,C,D\}$, and requested by $K=4$ users $\mathcal{K}=\{U_1,U_2,U_3,U_4\}$,
each has a cache of size $M=2$. Without loss of generality, we assume that each user request
file $A,B,C,D$ respectively.

 \begin{figure}
  \centering
  \includegraphics[width=4in]{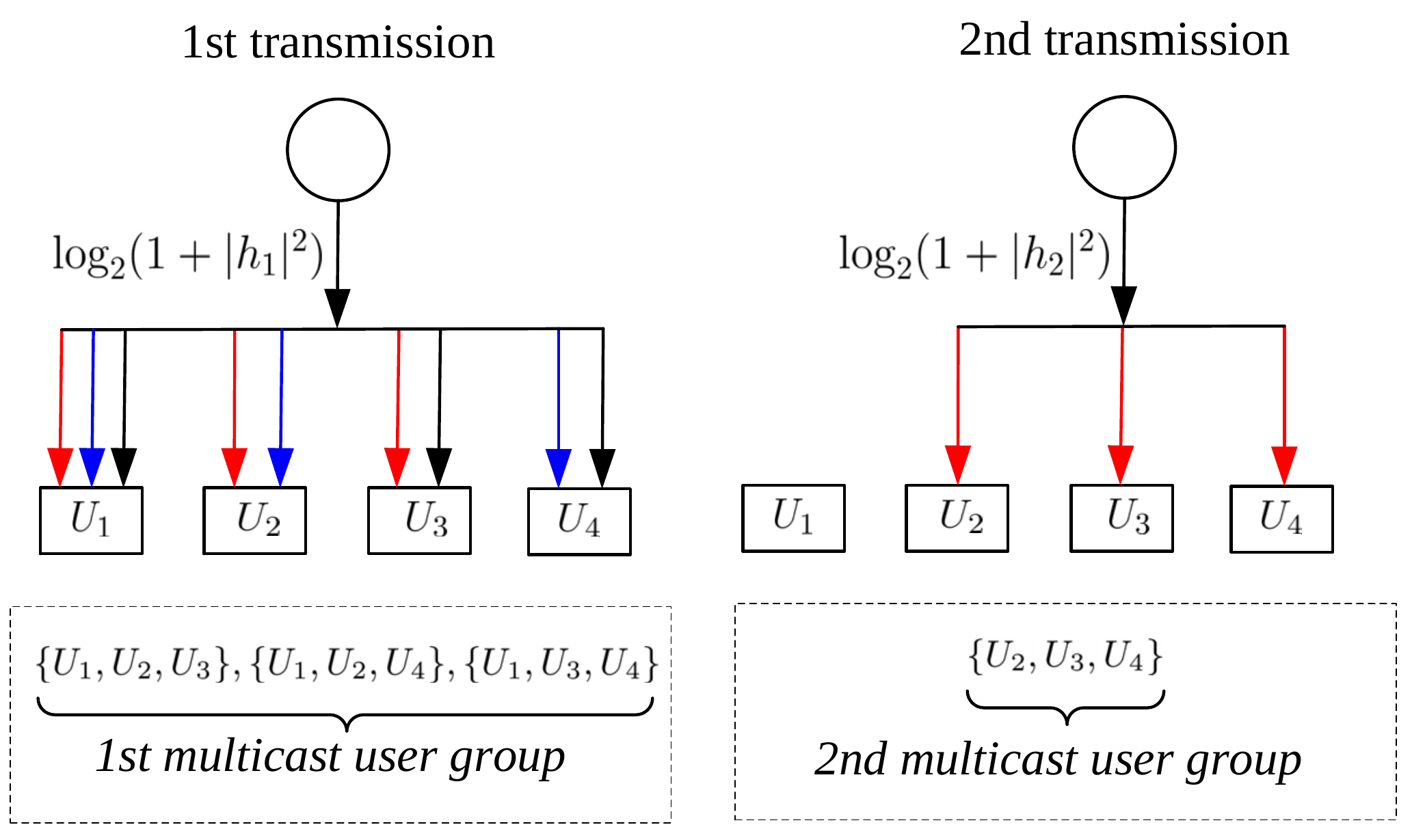}\\
  \caption{The procedures of orthogonal delivery scheme for centralized case.
  Where $N=K=4$ and $M=2$, the two multicast user group constrained
  by $U_1$ and $U_2$ are $\left\{\{U_1,U_2,U_3\},\{U_1,U_2,U_4\},\{U_1,U_3,U_4\}\right\}$ and $\{U_2,U_3,U_4\}$,
 transmitted with the maximum rate of $\log_2(1+|h_1|^2)$ and $\log_2(1+|h_2|^2)$ respectively.}\label{fig:centralizedo}
\end{figure}

Based on the Algorithm \ref{alg:centralizedplacement}, we have the file placement state:
\begin{align}
&Z_1=(W_{n,\{1,2\}},W_{n,\{1,3\}},W_{n,\{1,4\}})_{n=1}^4\\ \notag
&Z_2=(W_{n,\{1,2\}},W_{n,\{2,3\}},W_{n,\{2,4\}})_{n=1}^4\\ \notag
&Z_3=(W_{n,\{1,3\}},W_{n,\{2,3\}},W_{n,\{3,4\}})_{n=1}^4\\ \notag
&Z_4=(W_{n,\{1,4\}},W_{n,\{2,4\}},W_{n,\{3,4\}})_{n=1}^4
\end{align}
The multicast user group $\mathcal{G}_i (i=1,2)$ can be expressed as:
\begin{align}
&\mathcal{G}_1=\{\{U_1,U_2,U_3\},\{U_1,U_2,U_4\},\{U_1,U_3,U_4\}\}\\ \notag
&\mathcal{G}_2=\{\{U_2,U_3,U_4\}\}
\end{align}
Apparently, the transmission rate for $\mathcal{G}_1$ is constrained by channel capacity of $U_1$ , while it is constrained by
channel of $U_2$ for $\mathcal{G}_2$. Hence, the two delivery procedures are:
\begin{enumerate}
  \item send $\left\{A_{\{2,3\}}\oplus B_{\{1,3\}} \oplus C_{\{1,2\}};
A_{\{2,4\}}\oplus B_{\{1,4\}} \oplus C_{\{1,2\}};
A_{\{3,4\}}\oplus B_{\{1,4\}} \oplus C_{\{1,3\}}\right\}$ ~with rate $\log_2(1+|h_1|^2)$;
  \item send $\left\{A_{\{3,4\}}\oplus B_{\{2,4\}} \oplus C_{\{2,3\}}\right\}$ ~with rate $\log_2(1+|h_2|^2)$.
\end{enumerate}
To sum up, the total transmission time slots is:
\begin{align}
\mathcal{T}_{c,o}(\mathcal{H},2)=\frac{1}{2\log_2(1+|h_1|^2)}+\frac{1}{6\log_2(1+|h_2|^2)},
\end{align}
where $\mathcal{H}=(h_1,h_2)$.
\end{example}
\subsection{Concurrent Delivery Scheme}
For orthogonal delivery scheme, the multicast user group is served one by one,
during each transmission, the remaining channels stay in idle, which is bandwidth inefficiency.
While, with concurrent delivery scheme, the transmissions for all the multicast user groups are operated simultaneously, whereby, the power allocation
is essential to minimize the system transmission time slots which subjects to
the capacity region of GBCs.

Due to the same placement procedures with orthogonal delivery scheme, there are also $K-t$
multicast user groups, where $\binom{K-i}{t}$ multicast user sets are involved for each group $\mathcal{G}_i$.
However, the corresponding transmission rate $\mathbf{r}=(r_1,r_2,\cdots,r_{K-t})$ is decided by power allocation scheme,
and should satisfy $\mathcal{C}_{BC}$. At this time, the single server can be regarded as $(K-t)$ virtual sources to
transmit to separate multicast user group, with proper allocated rate respectively.
Similarly, the better users are able to decode the signal transmitted to the worse users additionally
and further to extract the requested segments of its own.

Obviously, the transmission time slots depend on the transmission rate $\mathbf{r}$, which
should satisfy the GBC capacity region shown as Lemma \ref{Cbc}.
And each $r_i, (i=1,\cdots,K-t)$ is constraint by the worst user in the associate group.
Thereby, we can formulate the transmission time slots as an optimization problem
to minimize the maximal transmission time slots for each group, which subjects to $\mathcal{C}_{BC}$:
\begin{align}\label{equ:optimizationorc}
&\min_{\textbf{r,s}} \quad \max \left\{\frac{\binom{K-1}{t}}{\binom{K}{t}r_1},\frac{\binom{K-2}{t}}{\binom{K}{t}r_2},\cdots,
\frac{\binom{K-(K-t)}{t}}{\binom{K}{t}r_{K-t}}\right\},\\ \notag
&~\mbox{s.t} \quad \quad\textbf{r}\in \mathcal{C}_{BC}.
\end{align}
Further, the problem in (\ref{equ:optimizationorc}) can be transformed into the following convex
optimization problem by using variable substitution:
\begin{align}\label{equ:convexopc}
&\min_\mathbf{r} \quad s \notag\\
&~\mbox{s.t}\quad \frac{\binom{K-i}{t}}{\binom{K}{t}r_i}\leq s; \notag\\
&\quad\quad\sum_{i=1}^{K-t-1} (\frac{1}{|h_i|^2}-\frac{1}{|h_{i+1}|^2})\times
2^{\sum \limits_{j=1}^i r_j}+\frac{1}{|h_{K-t}|^2}\times 2^{\sum \limits_{j=1}^{K-t} r_j}
-\frac{1}{|h_1|^2} \leq 1;\\ \notag
&\quad  \quad r_i\geq 0.\\ \notag
&\text{Where}\quad i=1,2,\cdots, K-t. \notag
\end{align}

\begin{lemma} \label{lemma:2}
For the convex optimization problem in (\ref{equ:convexopc}), the optimal value can be obtained at the boundary of the
inequalities.
\end{lemma}
The proof is shown in Appendix B.

According to Lemma \ref{lemma:2}, the broadcast capacity region in (\ref{equ:convexopc}) can be expressed and denoted by $g(s)$ as:
\begin{align}
g(s)=\sum_{i=1}^{K-t-1} (\frac{1}{|h_i|^2}-\frac{1}{|h_{i+1}|^2})\times
2^{\frac{\left(\binom{K}{t+1}-\binom{K-i}{t+1}\right)}{\binom{K}{t}s}}+\frac{1}{|h_{K-t}|^2}\times
2^{\frac{\binom{K}{t+1}}{\binom{K}{t}s}}-\frac{1}{|h_1|^2} -1=0.
\end{align}
Where $\sum \limits_{j=1}^{K-t} \binom{K-j}{t}=\binom{K}{t+1}\label{eq:com1},\sum \limits_{j=1}^i \binom{K-j}{t}=\binom{K}{t+1}-\binom{K-i}{t+1}$.

It is easy to observe that:
\begin{align}
&\lim \limits_{s\rightarrow +\infty} g(s)=-1,\\
&\lim \limits_{s\rightarrow 0^+} g(s)=+ \infty.
\end{align}
Namely, $g(s)$ is monotone decreasing in $(0,+ \infty)$.
Hence there exists a unique real root $s$ which corresponding to the transmission time slots of
centralized concurrent delivery scheme, denoted by
$\mathcal{T}_{c,p}(\mathcal{H},M)$.
\begin{theorem}
The transmission time slots $\mathcal{T}_{c,p}(\mathcal{H},M)$ of centralized
concurrent delivery scheme is the root of $g(s)$:\\
\begin{align}
g(s)=\sum \limits_{i=1}^{K-t-1} \left(\frac{1}{|h_i|^2}-\frac{1}{|h_{i+1}|^2}\right)\times
2^{\frac{\left(\binom{K}{t+1}-\binom{K-i}{t+1}\right)}{\binom{K}{t}s}}+\frac{1}{|h_{K-t}|^2}\times
2^{\frac{\binom{K}{t+1}}{\binom{K}{t}s}}-\frac{1}{|h_1|^2} -1=0.
\end{align}
\end{theorem}

For comparison with the orthogonal delivery scheme, the following algorithm describes
the concurrent delivery procedures of centralized case:
\begin{algorithm}
\caption{Centralized Delivery}\label{algorithm:para}
\begin{algorithmic}[1]
\Procedure {concurrent Delivery procedures}{}\\
$t\leftarrow KM/N$\\
Compute the $\mathcal{T}_{c,p}(\mathcal{H},M)$ from $g(s)=0$, with bisection method.
\For{$k=[K-t]$}\label{for}
\State $\mathcal{R}\leftarrow \left\{S=\{s_1,\cdots,s_{t+1}\},s_1=k\right\}$
\State $\mathbf{X}_k\leftarrow \oplus_{n\in S} W_{d_n,S\backslash\{n\}},S\in\mathcal{R}$
\EndFor\\
simultaneously send $\mathbf{X}_{\mathcal{D}}=\{\mathbf{X}_k\}$ with rate
$r_k=\frac{\binom{K-k}{t}}{\binom{K}{t}\mathcal{T}_{c,p}(\mathcal{H},M)}$ respectively.
\EndProcedure
\end{algorithmic}
\end{algorithm}

 \begin{figure}
  \centering
  \includegraphics[width=4in]{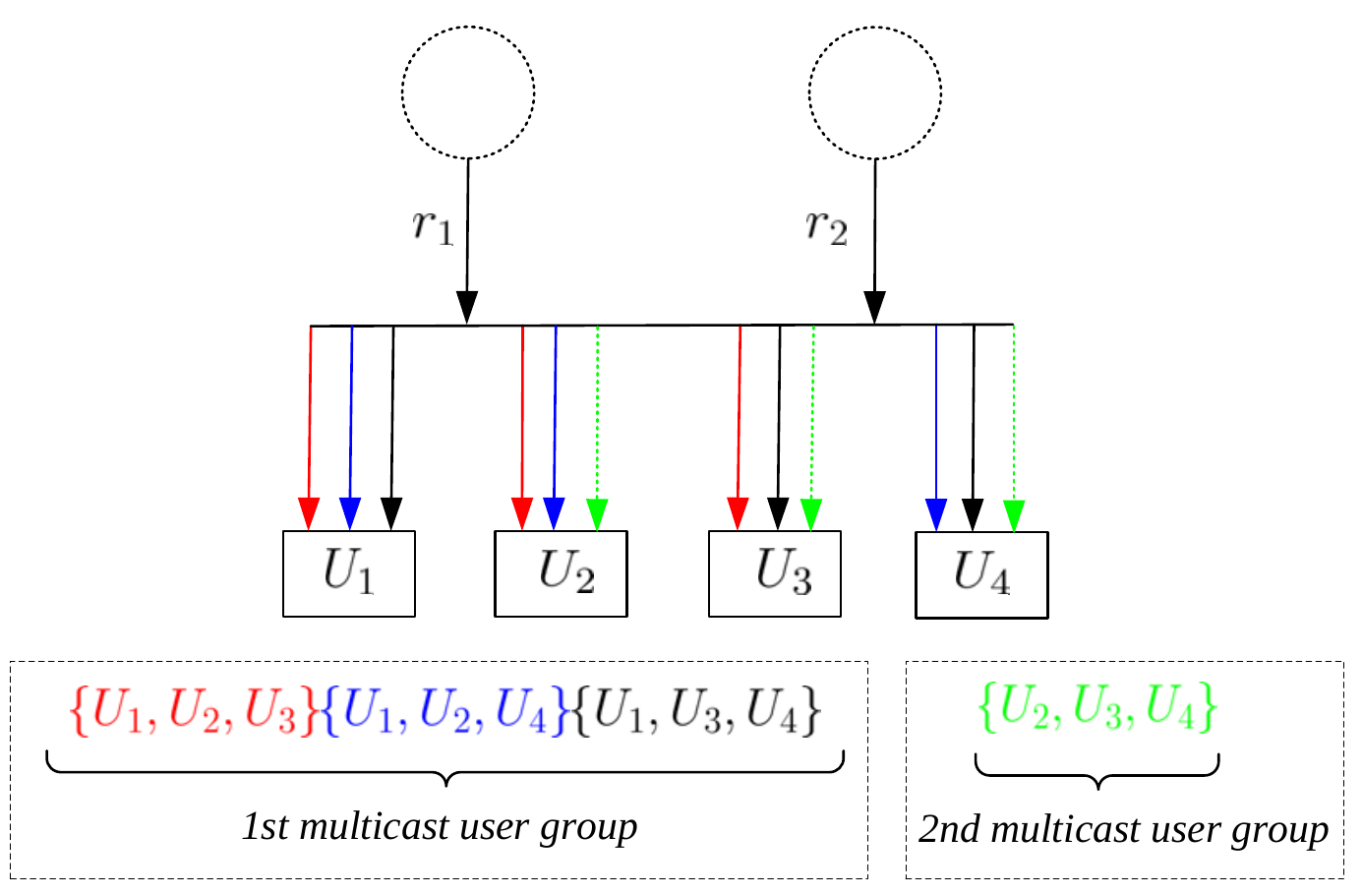}\\
  \caption{The procedures of concurrent delivery scheme of centralized case. $N=K=4, M=2$
  are assumed. The two multicast user groups are served with rate $r_1$ and $r_2$ respectively.}\label{fig:centralizedc}
\end{figure}

Where the server transmits the coded segment to all the $K-t$ multicast user groups simultaneously.
For completeness, we also take the above Example \ref{exa:ortde} for illustration, and the particular procedures of concurrent delivery scheme is shown as Fig. \ref{fig:centralizedc}.

\begin{example}
The difference only occurs during the delivery procedures.
For the concurrent delivery scheme, the server will concurrently send:
\begin{align}
\left\{
  \begin{array}{ll}
      \left\{\{A_{\{2,3\}}\oplus B_{\{1,3\}} \oplus C_{\{1,2\}};
A_{\{2,4\}}\oplus B_{\{1,4\}} \oplus C_{\{1,2\}};
A_{\{3,4\}}\oplus B_{\{1,4\}} \oplus C_{\{1,3\}}\}\right\},  \hbox{with rate $r_1=
\frac{1}{2\mathcal{T}_{c,p}(\mathcal{H},2)}$}; \\
    \left\{\{A_{\{3,4\}}\oplus B_{\{2,4\}} \oplus C_{\{2,3\}}\}\right\},  \hbox{with rate $r_2=
\frac{1}{6\mathcal{T}_{c,p}(\mathcal{H},2)}$.}
  \end{array}
\right.
\end{align}
where the concurrent operation is supported by using superposition code, and SIC is explored to extract the transmitted coded segments for each user. The system
transmission time slots turn out to be $\mathcal{T}_{c,p}(\mathcal{H},2)=\max\{\frac{1}{2r_1},\frac{1}{6r_2}\}$.
\end{example}

\section{Delivery Design for decentralized scheme}
Based on the knowledge from \cite{Ali2015}, the uppermost difference between centralized
and decentralized case is the placement procedures. For the decentralized case, the stored content
is chosen randomly and independently with each other.
For distinguishing with the centralized placement,
the Algorithm \ref{alg:deplace} firstly describes the procedures of decentralized placement:
\begin{algorithm}
\caption{Decentralized Placement}
\begin{algorithmic}[1]
\Procedure {Placement procedure}{}
\For{$k\in \mathcal{K}, W_n \in \mathcal{W}$}
\State user $k$ independently and uniformly caches a subset of $\frac{MF}{N}$ bits of file $W_n$.
\EndFor
\EndProcedure
\end{algorithmic}\label{alg:deplace}
\end{algorithm}

Similarly, the \emph{orthogonal delivery} and \emph{concurrent delivery} schemes can also be applied into the
decentralized case, and the transmission time slots will be assessed in the following subsections.
\subsection{Orthogonal delivery scheme}

Orthogonal delivery scheme for decentralized case is similar to the centralized case.
 The server send the coded segment to only one multicast user group per transmission
 with maximum rate, which constrained by the worst user in such group.
 The corresponding transmission time slots $\mathcal{T}_{d,o}(\mathcal{H},M)$ is given by Theorem \ref{the:tdo}.
\begin{theorem}\label{the:tdo}
The transmission time slots of decentralized orthogonal delivery scheme is:
\begin{align}
 \mathcal{T}_{d,o}(\mathcal{H},M)=\sum_{i=1}^K\frac{(1-q)^i}{\log_2(1+|h_i|^2)}.
 \end{align}
Where $q=\frac{M}{N}$.
\end{theorem}

Apparently, when transmit directly to the $i$th user, the series of users having better channel condition
can also receive and decode the signal correctly. Generally speaking, for user $i$, there are
in total $(1-q)$ non-stored files, however, due to the user $i$ has better channel condition than
the previous $i-1$ users and each bit is store by any user with the probability $q$,
the $\sum \limits_{j=1}^{i-1} q(1-q)^j$ files have already been sent
during the previous $i-1$ transmissions. That is, the server only need to send the remaining
$(1-q)^i$ files to the $i$th user with the allowed maximum
transmission rate $\log_2(1+|h_i|^2)$.

To give more insights into the above orthogonal delivery algorithm,
we take Example \ref{exa:deo} for illustration.
\begin{example} \label{exa:deo}
For a decentralized coded caching system, it is assumed that $K=N=2,M=1$,
the two files in library are denoted by $A,B$ and request by
$U_1$ and $U_2$ respectively. Each file is split
into $4$ segments denoted by $\mathcal{F}=\{F_k\},k=(0,1,2,12)$, where $F_k$ denotes the stored
segment by user $k$, especially, $F_0$ is the segment stored by neither of the two users.
Notablely, the size of each segment $F_k$ is $q^{|k|}(1-q)^{K-|k|}$, which reflects the
probability that each bit is store by $k$ users and missing in the remaining $K-|k|$ users,
whereby, $|\cdot|$ presents the element count operation.

Firstly, the placement state for two users are depicted as:
\begin{align}
Z_1=\{A_1,A_{12},B_1,B_{12}\},\\
Z_2=\{A_2,A_{12},B_2,B_{12}\}.\notag
\end{align}

Secondly, the
server will send:
\begin{enumerate}
  \item $\{A_0, A_2\oplus B_1\}$, with rate $\log_2(1+|h_1|^2)$;
  \item $\{B_0\}$, with rate $\log_2(1+|h_2|^2)$.
\end{enumerate}
\end{example}

Based on the two outputs and the respective placement states, $U_1$ and $U_2$ can extract the non-stored segments $A_0,A_2$
 and $B_0,B_1$ respectively.

Hence, the transmission time slots is sum up to:
\begin{align}
\mathcal{T}_{d,o}(\mathcal{H},1)=\frac{1}{2\log_2(1+|h_1|^2)}+\frac{1}{4\log_2(1+|h_2|^2)}.
\end{align}

\subsection{Concurrent delivery scheme}
The procedures of concurrent delivery scheme for decentralized placement tracks the centralized case.
However, the transmission load takes in difference, which resulted by the placement phase.
In the last subsection, we have proved that the transmission load directly for the
$i$th user is $(1-q)^i$, meanwhile, the signal is available for the next $K-i$ users having better channel
condition. Due to the concurrently operation, the transmission rate $\mathbf{r}=(r_1,\cdots,r_K)$ for distinct
users should satisfy $\mathcal{C}_{BC}$ as well. In the same way, the optimization problem is formulated:

\begin{align}\label{equ:decenop}
&\min_{\mathbf{r}}\quad \max\left\{\frac{(1-q)}{r_1},\frac{(1-q)^2}{r_2},\cdots,\frac{(1-q)^K}{r_K}\right\}, \notag\\
& ~\mbox{s.t} \quad \mathbf{r}\in \mathcal{C}_{BC}.
\end{align}

We further reformat (\ref{equ:decenop}) as a standard convex optimization problem:
\begin{align}
&\min_{\mathbf{r},t}\quad t\notag\\
&~\mbox{s.t}\quad \frac{(1-q)^k}{r_k}\leq t \notag\\
&\quad \quad \sum_{i=1}^{K-1}\left(\frac{1}{|h_i|^2}-\frac{1}{|h_{i+1}|^2}\right)\times
2^{\sum \limits_{j=1}^ir_j}+\frac{1}{|h_{K}|^2}\times
2^{\sum \limits_{j=1}^Kr_j}-\frac{1}{|h_1|^2}\leq1\label{const:capacity} \notag\\
&\quad \quad r_k\geq 0. \notag
\end{align}
Where $k\in\{1,2,\cdots,K\}$.
We have already shown in Appendix B that the optimal value can be obtained at the boundary
of the above subjects. Hence, for decentralized case, the transmission time slots of concurrent delivery scheme
$\mathcal{T}_{d,p}(\mathcal{H},M)$ can eventually be derived.
\begin{theorem}
The transmission time slot $\mathcal{T}_{d,p}(\mathcal{H},M)$ of decentralized
concurrent delivery scheme is the root of the following function:
\begin{align*}
g(t)=\sum_{i=1}^{K-1}\left(\frac{1}{|h_i|^2}-\frac{1}{|h_{i+1}|^2}\right)\times
2^{\frac{(1-q)\left(1-(1-q)^i\right)}{qt}} +\frac{1}{|h_{K}|^2}\times
2^{\frac{(1-q)\left(1-(1-q)^K\right)}{qt}}-\frac{1}{|h_1|^2}-1=0.
\end{align*}
\end{theorem}
Apparently, $g(t)$ is monotonic decreasing in $(0,+\infty)$. Hence,
there exists a unique real root which equals to $\mathcal{T}_{d,p}(\mathcal{H},M)$.

\section{Results for Designed Delivery Schemes}

In this section, we firstly compare the performance of the orthogonal delivery and concurrent
delivery scheme. The theoretical analysis results show that the concurrent delivery scheme is superior
both for centralized and decentralized placement cases. In addition, we also proved that when
adopting the same delivery scheme, i.e, orthogonal delivery scheme, the centralized placement case
takes shorter time in delivery phase, which is coincident with the conclusion carried out by Ali-Niesen considered with the error-free shared connected links,
and the gap between decentralized and centralized scheme is no more than $1.5$. It reveals that under the same conditions, centralized placement procedures
achieve more transmission efficiency.

The main results are listed as the following theorems.
\begin{theorem}\label{the:1}
The transmission time slots of orthogonal and concurrent delivery schemes for centralized placement case satisfy:
\begin{align}
\mathcal{T}_{c,p}(\mathcal{H},M)\leq \mathcal{T}_{c,o}(\mathcal{H},M).
\end{align}
\end{theorem}
\begin{theorem}\label{the:2}
The transmission time slots of orthogonal and concurrent delivery schemes for decentralized placement case satisfy:
\begin{align}
\mathcal{T}_{d,p}(\mathcal{H},M)\leq \mathcal{T}_{d,o}(\mathcal{H},M).
\end{align}
\end{theorem}
Both for Theorem \ref{the:1} and Theorem \ref{the:2}, when and only
when $|h_1|=|h_2|=\cdots=|h_K|$, the equation is established.
\begin{theorem}\label{the:3}
When adopting the same orthogonal deliver scheme, the gap between the decentralized and
centralized case is bounded by 1.5, which is:
\begin{align}
 \frac{\mathcal{T}_{d,o(\mathcal{H},M)}}{\mathcal{T}_{c,o(\mathcal{H},M)}} \leq 1.5
\end{align}
In special, when $K=2,t=1$, the gap between decentralized and centralized cases is the deepest to 1.5.
\end{theorem}
The proof of the Theorem \ref{the:1}, \ref{the:2}, \ref{the:3} are shown
in Appendix C, D, E respectively.

\begin{figure}
  \centering
  \includegraphics[width=3.5in]{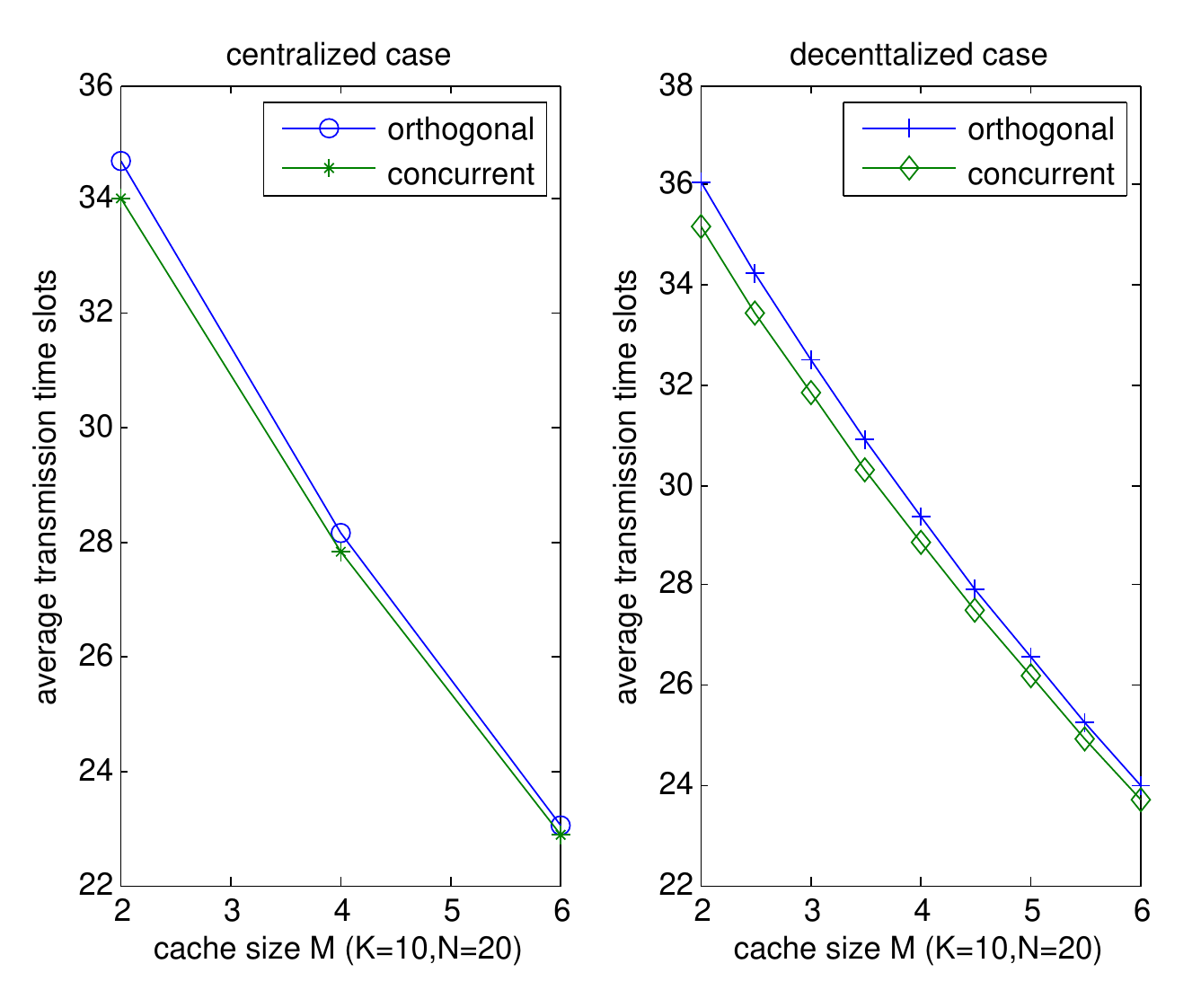}\\
  \caption{The average transmission time slots of orthogonal delivery and concurrent delivery scheme for centralized (left)
  and decentralized (right) cases.}\label{fig:transmission_time}
\end{figure}

To validate our analysis results, we simulate the average transmission time slots vary with the cache
 size $M$ of
our proposed orthogonal delivery and concurrent delivery scheme. As shown in Fig. \ref{fig:transmission_time},
either for centralized or decentralized case, the concurrent delivery scheme outperforms orthogonal delivery scheme,
this superiority comes from the involved power allocation strategy for concurrent deliver scheme,
which guarantees to make utmost use of all the channels. Meanwhile, it reveals that with the cache size increasing,
the advantage of concurrent delivery scheme is gradually vanishing.
When respectively considering the centralized or decentralized case,
the transmission load
stays in the same, while, the results show that the difference of transmission time slots for different delivery scheme is not vast, which indicates
that the system performance achieved by delivery scheme is less significant than directly reducing transmission load.
In addition, it also shows that under the same delivery scheme, the average transmission time slots of centralized case is much shorter than decentralized case.

\begin{figure}
  \centering
  \includegraphics[width=3in]{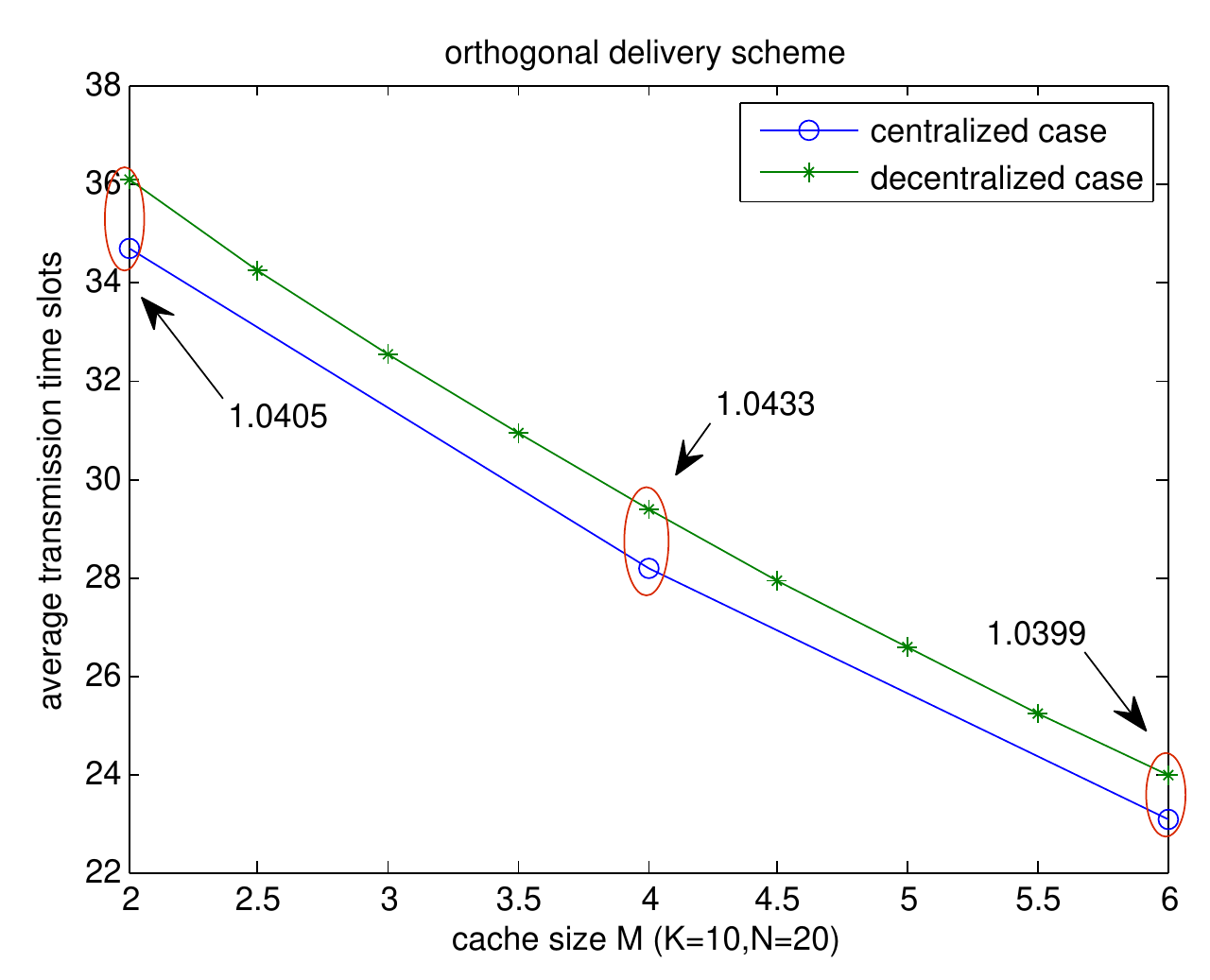}\\
  \caption{When adopting the same orthogonal delivery scheme, the
  average transmission time slots and the gap between decentralized an centralized placement
  vary with the cache size $M$.}\label{fig:orthgonal_gap}
\end{figure}

Then, in Fig. \ref{fig:orthgonal_gap}, we further assess the average transmission time slots influenced by different
placement procedures, to keep the comparison conditions remain in the same, we take the same orthogonal delivery scheme into
consideration, it shows that under the same delivery procedures, the transmission
time slots of centralized case is much more shorter, this gain benefits from the carefully designed placement, which
creates more coded multicasting opportunities. Additionally, the gap of our simulation result effectively validates our
theoretic analysis, which is no more than 1.5.

\section{Conclusion}
In this paper, we explore the coded caching scheme into Gaussian Broadcast Networks both for centralized
and decentralized cases. Firstly, for both placement cases, we proposed two delivery schemes, namely:
\emph{orthogonal delivery} and \emph{concurrent delivery} scheme. Then, the transmission time slots with different proposed delivery schemes is assessed, which
indicates the transmission delay when compared with the coded caching over error-free shared links.
Further, we proved the concurrent delivery scheme cost shorter time in delivery phase within
the same placement procedures, i.e., centralized and decentralized placement. Finally, the average transmission
time slots are also compared between different placement procedures, where the orthogonal delivery scheme is adopted,
the gap of transmission time slots between decentralized and centralized case is turned out to be no more than 1.5,
which coincides with the gap under error-free and shared links.

Even though, the concurrent delivery scheme is turned out to be a strategy to reduce the system transmission time slots,
while, for such users having better channel conditions, the individual delay is expanded. In our future work,
we expect to investigate another opportunistic multicast scheme to jointly considering the utilization efficiency of
better channels.

\appendices
\section{Proof of the Lemma \ref{Cbc}}
Consider a more general network with bandwidth $W$, transmit power and noise power spectral density
are $P$ and $\frac{N_0}{2}$ respectively, and it is assumed the channel gain is increasing with the index
of users, i.e., $|h_1|\leq |h_2|\leq \cdots\leq |h_K|$, $|h_{K+1}|=0$ for special. The original
Gaussian Broadcast Channel capacity region is:
\begin{align}
\mathcal{C}_{BC}^{\star}=\left\{\mathbf{r}~\left|~ r_i\leq W \log_2\left(1+\frac{\alpha_i |h_i|^2 P}
{N_0 W+\sum \limits_{j=i+1}^K \alpha_j P |h_i|^2}\right), 0\leq \alpha_i \leq 1, \sum \limits_{i=1}^K
\alpha_i =1, r_i \geq 0, \forall i \right. \right\}.\label{equ:bc1}
\end{align}
While the equivalent capacity region is presented by:
\begin{equation}
\begin{aligned}
\mathcal{C}_{BC}=\left\{\mathbf{r}~\left| \sum \limits_{i=1}^K \left(\frac{1}{|h_i|^2}- \frac{1}{|h_{i+1}|^2}\right)
2^{\sum \limits_{j=1}^i\frac{r_j}{W}}- \frac{1}{|h_1|^2}\leq \frac{P}{N_0W},r_i\geq0, \forall i,\frac{1}{|h_{K+1}|^2}=0\right. \right\}.
\end{aligned}
\end{equation}
Firstly, we prove $\mathcal{C}_{BC}^{\star}\subset \mathcal{C}_{BC}$, and then
$\mathcal{C}_{BC}\subset \mathcal{C}_{BC}^{\star}$.
\subsection{The proof of $\mathcal{C}_{BC}^{\star}\subset \mathcal{C}_{BC}$.}
The above problem is changed to prove the satisfaction of the following inequality
for any $0< m\leq K$,
under the condition that $\mathbf{r}\in \mathcal{C}_{BC}^{\star}$ by using mathematical induction:
\begin{align}
\left(\sum \limits_{i=K-m+1}^K \alpha_i \right)\frac{P}{N_0W}\geq
\sum \limits_{i=K-m+1}^K \left(\frac{1}{|h_i|^2}-\frac{1}{|h_{i+1}|^2}\right)
2^{\sum \limits_{j=K-m+1}^i \frac{r_j}{W}}-\frac{1}{|h_{K-m+1}|^2}.\label{equ:pro1}
\end{align}
\begin{enumerate}
  \item $m=1$. 
\begin{align}
r_K\leq W\log_2\left(1+\frac{\alpha_K |h_K|^2 P}{N_0W}\right)
\Rightarrow\left(2^{\frac{r_K}{W}}-1\right)\frac{1}{|h_K|^2}\leq \frac{\alpha_K P}{N_0W}.\notag
\end{align}

  \item $m=n, n<K$. It is assumed that (\ref{equ:pro1}) is satisfied, that is to say for any $m=n, \left(n<K\right)$, we have
\begin{align}
\left(\sum \limits_{i=K-n+1}^K \alpha_i \right)\frac{P}{N_0W}\geq
\sum \limits_{i=K-n+1}^K \left(\frac{1}{|h_i|^2}-\frac{1}{|h_{i+1}|^2}\right)
2^{\sum \limits_{j=K-n+1}^i \frac{r_j}{W}}-\frac{1}{|h_{K-n+1}|^2}\label{equ:jiashe1}.
\end{align}
  \item $m=n+1$. Due to the condition that $\mathbf{r}\in \mathcal{C}_{BC}^{\star}$, we have
  \begin{align}
r_{K-n}&\leq W\log_2\left(1+\frac{\alpha_{K-n}|h_{K-n}|^2 P}{N_0W+\sum \limits_{j={K-n+1}}^K \alpha_j |h_{K-n}|^2 P}\right),
\end{align}
and
\begin{align}
\left(\sum \limits_{j=K-n}^K \alpha_j\right) \frac{P}{N_0W}\geq
2^{\frac{r_{K-n}}{W}}\left(\frac{1}{|h_{K-n}|^2}+\frac{P}{N_0W}\sum \limits_{j={K-n+1}}^K \alpha_j\right)-\frac{1}{|h_{K-n}|^2}.\label{equ:alp}
\end{align}
Plug (\ref{equ:jiashe1}) into (\ref{equ:alp}), we may derive
\begin{align}
\left(\sum \limits_{j=K-n}^K \alpha_j\right) \frac{P}{N_0W}\geq
\sum \limits_{i=K-n}^K \left(\frac{1}{|h_i|^2}-\frac{1}{|h_{i+1}|^2}\right)2^{\sum \limits_{j=K-n}^i \frac{r_j}{W}}-\frac{1}{|h_{K-n}|^2}.
\end{align}
\end{enumerate}
It is shown that, for any given $0< m\leq K$, (\ref{equ:pro1}) is always satisfied. Specially, when $m=K$:
\begin{align}
\left(\sum \limits_{i=1}^K \alpha_i \right)\frac{P}{N_0W}\geq
\sum \limits_{i=1}^K \left(\frac{1}{|h_i|^2}-\frac{1}{|h_{i+1}|^2}\right)
2^{\sum \limits_{j=1}^i \frac{r_j}{W}}-\frac{1}{|h_{1}|^2}\label{equ:pro2}.
\end{align}
Since $\sum \limits_{i=1}^K \alpha_i =1$, consequently, we have
\begin{align}
\frac{P}{N_0W}\geq
\sum \limits_{i=1}^K \left(\frac{1}{|h_i|^2}-\frac{1}{|h_{i+1}|^2}\right)
2^{\sum \limits_{j=1}^i \frac{r_j}{W}}-\frac{1}{|h_{1}|^2}.\label{equ:pro2}
\end{align}
Obviously, all the above analysis leads to $\mathcal{C}_{BC}^{\star}\subset \mathcal{C}_{BC}$.
\subsection{The proof of $\mathcal{C}_{BC}\subset \mathcal{C}_{BC}^{\star}$.}
On the contrary, we need to prove $\mathbf{r}\in \mathcal{C}_{BC}^{\star}$ if $\mathbf{r}\in \mathcal{C}_{BC}$. In other words, we need to prove that there
exist $0\leq \alpha_i\leq 1,\sum \limits_{i=1}^K \alpha_i=1$ and
\begin{align}
r_i\leq W\log_2\left(1+\frac{\alpha_i |h_i|^2 P}{N_0W+\sum \limits_{j>i}\alpha_j P|h_i|^2 }\right).
\end{align}
Let
\begin{align}
&\delta_{K-m+1}=\frac{N_0W}{P}\left(\sum \limits_{i=K-m+1}^K \left(\frac{1}{|h_i|^2}-\frac{1}{|h_{i+1}|^2}\right)2^{\sum \limits_{j=K-m+1}^i \frac{r_j}{W}}-\frac{1}{|h_{K-m+1}|^2}\right)\leq 1,\\
&\delta_{K-m}=\frac{N_0W}{P}\left(\sum \limits_{i=K-m}^K \left(\frac{1}{|h_i|^2}-\frac{1}{|h_{i+1}|^2}\right)2^{\sum \limits_{j=K-m}^i \frac{r_j}{W}}-\frac{1}{|h_{K-m}|^2}\right)\leq 1,
\end{align}
where $\delta_{K-m}\geq \delta_{K-m+1},\delta_1\leq 1$. Obviously we have
\begin{align}
\delta_{K-m}=\frac{N_0W}{P}\left(2^{\frac{r_{K-m}}{W}}\left(\frac{P}{N_0W}\delta_{K-m+1}+\frac{1}{|h_{K-m}|^2}\right)-
\frac{1}{|h_{K-m}|^2}\right).\label{equ:pro3}
\end{align}
Then $r_{K-m}$ can be calculated as below
\begin{align}
r_{K-m}=W\log_2\left(1+\frac{\left(\delta_{K-m}-\delta_{K-m+1}\right)P|h_{K-m}|^2}
{N_0W+\delta_{K-m+1}P|h_{K-m}|^2}\right).\label{equ:pro4}
\end{align}
Further let $\alpha_i=\delta_i-\delta_{i+1},\forall i, i\in\{2,3,\cdots,K\},\delta_{K+1}=0,\alpha_1=1-\delta_2$. It is easy to derive that
$\alpha_i\geq 0,\sum \limits_{i=1}^K \alpha_i =1$. Then, we take $i=K-m$ into (\ref{equ:pro4}) to have
\begin{align}
r_i=W\log_2\left(1+\frac{\alpha_i P |h_i|^2}{N_0W+\delta_{i+1} P |h_i|^2}\right)
=W\log_2\left(1+\frac{\alpha_i P |h_i|^2}{N_0W+\sum \limits_{j>i}\alpha_j P |h_i|^2}\right).
\end{align}
When $i=1$
\begin{align}
r_1&=W\log_2\left(1+\frac{\left(\delta_{1}-\delta_{2}\right) P |h_1|^2}{N_0W+\sum \limits_{j>1}\alpha_j P |h_1|^2}\right)\\ \notag
&\leq
W\log_2\left(1+\frac{\alpha_1 P |h_1|^2}{N_0W+\sum \limits_{j>1}\alpha_j P |h_1|^2}\right).\notag
\end{align}
Therefore, the region $\mathcal{C}_{BC}\subset \mathcal{C}_{BC}^{\star}$ based
on the condition $\mathbf{r}\in \mathcal{C}_{BC}$ can be achieved.

Combining the analysis in subsections \emph{A} and \emph{B}, the two regions of $\mathcal{C}_{BC}^{\star}$ and $\mathcal{C}_{BC}$
are proved to be equivalent.

\section{proof of the Lemma \ref{lemma:2}}
The dual function of the convex optimization problem in (\ref{equ:convexopc}) can be expressed as
\begin{align}
&\mathcal{L}(s,r_i,\alpha_i,\lambda)=
&s+\sum \limits_{i=1}^{K-t} \alpha_i \left(\frac{\binom{K-i}{t}}{\binom{K}{t}r_i}-s
\right)+\lambda \left(\sum_{i=1}^{K-t-1}\left(\frac{1}{|h_i|^2}-\frac{1}{|h_{i+1}|^2}\right)2^{\sum \limits_{j=1}^ir_j}+\frac{1}{|h_{K-t}|^2}2^{\sum \limits_{j=1}^{K-t}r_j}-\frac{1}{|h_1|^2}-1\right),\notag
\end{align}
where $i=1,\cdots,K-t$. We may derive the following KKT conditions.
\begin{align}
\left\{
  \begin{array}{ll}
    \alpha_i\geq 0,  \\
     \lambda\geq 0,  \\
     \frac{\partial \mathcal{L}(s,r_i,\alpha_i,\lambda)}{\partial s}=0,  \\
    \frac{\partial \mathcal{L}(s,r_i,\alpha_i,\lambda)}{\partial r_i}=0,  \\
    \alpha_i \left(\frac{\binom{K-i}{t}}{\binom{K}{t}r_i}-s\right)=0,  \\
    \lambda \left(\sum \limits_{i=1}^{K-t-1}\left(\frac{1}{|h_i|^2}-\frac{1}{|h_{i+1}|^2}\right)2^{\sum \limits_{j=1}^ir_j}+\frac{1}{|h_{K-t}|^2}2^{\sum \limits_{j=1}^{K-t}r_j}-\frac{1}{|h_1|^2}-1\right)=0,\\
    i=1,2,\cdots,K-t.
  \end{array}
\right.
\end{align}
When taking the derivation of $s$ and $r_i$, if we set $\lambda=0$, it results that $\alpha_i=0$, while conflicts with the condition that $\frac{\partial \mathcal{L}(s,r_i,\alpha_i,\lambda)}{\partial s}=1-\sum \limits_{i=1}^{K-t} \alpha_i=0$, Hence $\lambda>0, \alpha_i>0, (i=1,\cdots,K-t)$.
Moreover, $\lambda, \alpha_i$ should satisfy the following equality constraints

\begin{align}
\left\{
  \begin{array}{ll}
    \alpha_i \left(\frac{\binom{K-i}{t}}{\binom{K}{t}r_i}-s\right)=0,\\
    \lambda \left(\sum \limits_{i=1}^{K-t-1}\left(\frac{1}{|h_i|^2}-\frac{1}{|h_{i+1}|^2}\right)2^{\sum \limits_{j=1}^ir_j}+\frac{1}{|h_{K-t}|^2}2^{\sum \limits_{j=1}^{K-t}r_j}-\frac{1}{|h_1|^2}-1\right)=0.
  \end{array}
\right.
\end{align}
Hence, we have
\begin{align}
\left\{
  \begin{array}{ll}
\frac{\binom{K-i}{t}}{\binom{K}{t}r_i}-s=0, \\
\sum \limits_{i=1}^{K-t-1}\left(\frac{1}{|h_i|^2}-\frac{1}{|h_{i+1}|^2}\right)2^{\sum \limits_{j=1}^ir_j}+\frac{1}{|h_{K-t}|^2}2^{\sum \limits_{j=1}^{K-t}r_j}-\frac{1}{|h_1|^2}-1=0.
  \end{array}
\right.
\end{align}
And this implies that the optimal value can be obtained at the boundary point.

\section{The proof of Theorem \ref{the:1}}
In order to simplify the expression, we use $\mathcal{T}_{\star}$ to replace $\mathcal{T}_{\star}(\mathcal{H},M)$. Since $g(s)$ is monotone decreasing in $(0,+\infty)$, and $\mathcal{T}_{c,p}$ is the root of $g(s)$, we only need to prove $g(\mathcal{T}_{c,o})\leq g(\mathcal{T}_{c,p})=0$.

Firstly, when $|h_1|=|h_2|=\cdots=|h_K|$,
\begin{align}
&g(\mathcal{T}_{c,o})=0.
\end{align}
It indicates that $\mathcal{T}_{c,o}=\mathcal{T}_{c,p}$.

Secondly, when $|h_1|<|h_2|<\cdots<|h_K|$, the mathematical induction is adopted. Here, we aim to prove that for any arbitrary $|h_1|<|h_2|<\cdots<|h_K|; 0<t<K; t,K \in \mathbb{N}^+$, there exists:
\begin{align}\label{eq:cenp0}
g(\mathcal{T}_{c,o})=\sum_{i=1}^{K-t-1} \left(\frac{1}{|h_i|^2}-\frac{1}{|h_{i+1}|^2}\right)\times 2^{\frac{\left(\binom{K}{t+1}-\binom{K-i}{t+1}\right)}{\sum \limits_{l=1}^{K-t}\frac{\binom{K-l}{t}}{\log_2(1+|h_l|^2)}}}+\frac{1}{|h_{K-t}|^2}\times 2^{\frac{\binom{K}{t+1}}{\sum \limits_{l=1}^{K-t}\frac{\binom{K-l}{t}}{\log_2(1+|h_l|^2)}}}-\frac{1}{|h_1|^2} -1\leq 0.
\end{align}
step 1:

When $K=2,t=1, \forall |h_i|\in \mathbb{C},|h_i|>0$, when taken into (\ref{eq:cenp0}), we have
\begin{align}
g(\mathcal{T}_{c,o})=\frac{1}{|h_1|^2}\times 2^{\log_2(1+|h_1|^2)}-\frac{1}{|h_1|^2} -1=0.
\end{align}
step 2:

It is assumed that the inequality (\ref{eq:cenp0}) is satisfied for arbitrary $K=n; 0<t<n;t,n\in\mathbb{N}^+$, where $\mathcal{H}=(|h_2|,|h_3|,\cdots,|h_{n+1-t}|)$ and $|h_2|\leq|h_3|\leq\cdots\leq|h_{n+1-t}|$. Then we have
\begin{align}\label{eq:cenp1}
&g(\mathcal{T}_{c,o})=
\sum_{j=1}^{n-t-1} \left(\frac{1}{|h_{j+1}|^2}-\frac{1}{|h_{j+2}|^2}\right)\times 2^{\frac{\left(\binom{n}{t+1}-\binom{n-j}{t+1}\right)}{\sum \limits_{l=1}^{n-t}\frac{\binom{n-l}{t}}{\log_2(1+|h_{l+1}|^2)}}}+\frac{1}{|h_{n+1-t}|^2}\times 2^{\frac{\binom{n}{t+1}}{\sum \limits_{l=1}^{n-t}\frac{\binom{n-l}{t}}{\log_2(1+|h_{l+1}|^2)}}}-\frac{1}{|h_2|^2} -1\leq 0.
\end{align}
Let us set
\begin{align}
&b_j=\binom{n}{t+1}-\binom{n-j}{t+1},\\
&b_{n-t}=\binom{n}{t+1},\\
&a=\sum \limits_{l=1}^{n-t}\frac{\binom{n-l}{t}}{\log_2(1+|h_{l+1}|^2)}.
\end{align}
Then the inequality (\ref{eq:cenp1}) can be rewritten as
\begin{align}
g(\mathcal{T}_{c,o})=\sum_{j=1}^{n-t-1} (\frac{1}{|h_{j+1}|^2}-\frac{1}{|h_{j+2}|^2})\times 2^{\frac{b_j}{a}}+\frac{1}{|h_{n+1-t}|^2}\times 2^{\frac{b_{n-t}}{a}}-\frac{1}{|h_2|^2} -1\leq 0.\label{eq:jiashe}
\end{align}

step 3: Further, we need to prove for $K=n+1$, the inequality (\ref{eq:jiashe}) is still satisfied. Now $g(\mathcal{T}_{c,o})$ is given by
\begin{align}\label{eq:nplus1}
g(T_{c,o})=\sum_{i=1}^{n-t} (\frac{1}{|h_{i}|^2}-\frac{1}{|h_{i+1}|^2})\times 2^{\frac{\left(\binom{n+1}{t+1}-\binom{n+1-i}{t+1}\right)}{\sum \limits_{l=1}^{n+1-t}\frac{\binom{n+1-l}{t}}{\log_2(1+|h_{l}|^2)}}}+\frac{1}{|h_{n+1-t}|^2}\times 2^{\frac{\binom{n+1}{t+1}}{\sum \limits_{l=1}^{n+1-t}\frac{\binom{n+1-l}{t}}{\log_2(1+|h_{l}|^2)}}}-\frac{1}{|h_1|^2} -1,
\end{align}
where we set:
\begin{align}
&c_i=\binom{n+1}{t+1}-\binom{n+1-i}{t+1},\label{equ:ci}\\
&c_{n+1-t}=\binom{n+1}{t+1}, \\
&d=\sum \limits_{l=1}^{n+1-t}\frac{\binom{n+1-l}{t}}{\log_2(1+|h_{l}|^2)}.
\end{align}
For $c_i$, let $j=i-1$, we can derive that
\begin{align}\label{equ:cj}
c_{j+1}
=\delta'+b_j,
\end{align}
Where $\delta'=\binom{n}{t}$.
Moreover, for $d$:
\begin{align}\label{equ:d2}
d
=\delta+a,
\end{align}
Where $\delta=\frac{\binom{n}{t}}{\log_2(1+|h_1|^2)}$.
Substituting from (\ref{equ:ci}) to (\ref{equ:d2}) into (\ref{eq:nplus1}), we have
\begin{align}
g(\mathcal{T}_{c,o})=&\sum_{i=1}^{n-t} (\frac{1}{|h_{i}|^2}-\frac{1}{|h_{i+1}|^2})\times 2^{\frac{c_i}{d}}+\frac{1}{|h_{n+1-t}|^2}\times 2^{\frac{c_{n+1-t}}{d}}-\frac{1}{|h_1|^2} -1 \notag\\
=&(\frac{1}{|h_{1}|^2}-\frac{1}{|h_{2}|^2})\times 2^{\frac{c_1}{d}}+\sum_{i=2}^{n+1-t-1} (\frac{1}{|h_{i}|^2}-\frac{1}{|h_{i+1}|^2})\times 2^{\frac{c_i}{d}}+\frac{1}{|h_{n+1-t}|^2}\times 2^{\frac{c_{n+1-t}}{d}}-\frac{1}{|h_1|^2} -1 \notag\\
\overset{j=i-1}{=}
&(\frac{1}{|h_{1}|^2}-\frac{1}{|h_{2}|^2})\times 2^{\frac{c_1}{d}}+\sum_{j=1}^{n-t-1} (\frac{1}{|h_{j+1}|^2}-\frac{1}{|h_{j+2}|^2})\times 2^{\frac{b_j+\delta'}{a+\delta}}+\frac{1}{|h_{n+1-t}|^2}\times 2^{\frac{c_{n+1-t}}{d}}-\frac{1}{|h_1|^2} -1.\label{eq:main}
\end{align}
According to Jessen inequality, we have
\begin{align}\label{eq:inequal1}
&2^{\frac{c_1}{d}}=2^{\frac{\delta'}{a+\delta}}=2^{\frac{a}{a+\delta}\times\frac{0}{a}+
\frac{\delta}{a+\delta}\frac{\delta'}{\delta}}
\leq \frac{a}{a+\delta}+\frac{\delta}{a+\delta}(1+|h_1|^2),\\
&2^{\frac{b_j+\delta'}{a+\delta}}=2^{\frac{a}{a+\delta}\times\frac{b_j}{a}
+\frac{\delta}{a+\delta}\frac{\delta'}{\delta}}
\leq \frac{a}{a+\delta}2^{\frac{b_j}{a}}+\frac{\delta}{a+\delta}(1+|h_1|^2),\\
&2^{\frac{c_{n+1-t}}{d}}=2^{\frac{b_{n-t}+\delta'}{a+\delta}}=2^{\frac{a}{a+\delta}\times\frac{b_{n-t}}{a}
+\frac{\delta}{a+\delta}\frac{\delta'}{\delta}}
\leq \frac{a}{a+\delta}2^{\frac{b_{n-t}}{a}}+\frac{\delta}{a+\delta}(1+|h_1|^2).\label{eq:inequal3}
\end{align}
Now we have
\begin{align}
g(\mathcal{T}_{c,o}) \leq
&
(\frac{1}{|h_{1}|^2}-\frac{1}{|h_{2}|^2})\times \left(\frac{a}{a+\delta}+\frac{\delta}{a+\delta}(1+|h_1|^2)\right) \notag\\
&+\frac{a}{a+\delta}\left\{\sum_{j=1}^{n-t-1} (\frac{1}{|h_{j+1}|^2}-\frac{1}{|h_{j+2}|^2})\times 2^{\frac{b_j}{a}}+\frac{1}{|h_{n+1-t}|^2}\times
2^{\frac{b_{n-t}}{a}}\right\} \notag\\
&+\frac{\delta}{a+\delta}\left\{(\frac{1}{|h_{2}|^2}-\frac{1}{|h_{n+1-t}|^2})(1+\frac{1}{|h_1|^2})
+\frac{1}{|h_{n+1-t}|^2}(1+\frac{1}{|h_1|^2})\right\}-\frac{1}{|h_1|^2} -1 \notag\\
\leq
&(\frac{1}{|h_{1}|^2}-\frac{1}{|h_{2}|^2})\times (\frac{a}{a+\delta}+\frac{\delta}{a+\delta}(1+|h_1|^2))+\frac{a}{a+\delta}(\frac{1}{|h_2|^2} +1) \notag\\
&+\frac{\delta}{a+\delta}(\frac{1}{|h_2|^2}(1+|h_1|^2))-\frac{1}{|h_1|^2} -1 \notag\\
=&0.
\label{eq:main2}
\end{align}

step 4:
In the previous proof, we assumed that $t<K$, while for $K=n+1$, the case $t=n$ has not been contained. Additionally, for $K=n+1, t=n$, we have:
\begin{align}
g(\mathcal{T}_{c,o})=\frac{1}{|h_1|^2}\times 2^{\frac{1}{\frac{1}{\log_2(1+|h_1|^2)}}}-\frac{1}{|h_1|^2}-1
=0
\end{align}

Consequently, we have proved that $g(\mathcal{T}_{c,o})\leq 0$ which reveals that $\mathcal{T}_{c,p}\leq \mathcal{T}_{c,o}$ for any arbitrary $|h_1|\leq |h_2|\leq \cdots \leq |h_{K-t}|$.

\section{The proof of Theorem \ref{the:2}}
By using the similar mathematical induction as in the Appendix C, we may prove $\mathcal{T}_{d,p}\leq \mathcal{T}_{d,o}$.

Firstly, when $|h_1|=|h_2|=\cdots=|h_K|$. It is easy to derive that $g(\mathcal{T}_{d,o})=g(\mathcal{T}_{d,p})=0$, hence, $\mathcal{T}_{d,p}=\mathcal{T}_{d,o}$ holds.

Secondly, when $|h_1|<|h_2|<\cdots<|h_K|$, the mathematical induction is involved into proof.

Step 1:
when $K=1$, it is easy to derive that $g(\mathcal{T}_{d,o})=0$, which indicates that $\mathcal{T}_{d,o}=\mathcal{T}_{d,p}$.

Step 2:
when $K>1$, it is assumed $g(t)\leq 0$ is also satisfied, that is to say:
\begin{align}
g(t)=\sum_{i=1}^{K-1}\left(\frac{1}{|h_i|^2}-\frac{1}{|h_{i+1}|^2}\right)2^{\frac{b_i}{a}}+\frac{1}{|h_K|^2}2^{\frac{b_K}{a}}-\frac{1}{|h_1|^2}-1\leq 0,\label{neq1}
\end{align}
for any $|h_1|<|h_2|<\cdots<|h_K|$, where $a=\sum\limits_{i=1}^K\frac{(1-q)^i}{\log_2(1+|h_i|^2)}$, $b_i=\frac{(1-q)\left(1-(1-q)^i\right)}{q}$.

Step 3:
Considering that there are $K+1$ users, and $|h_1|<|h_2|<\cdots<|h_K|<|h_{K+1}|$, let $\delta=\frac{(1-q)^{K+1}}{\log_2(1+|h_{K+1}|^2)}$, we have
\begin{align}
g\left(\mathcal{T}_{d,o}\right)=\sum_{i=1}^{K}\left(\frac{1}{|h_i|^2}-\frac{1}{|h_{i+1}|^2}\right)2^{\frac{b_i}{a+\delta}}+\frac{1}{|h_{K+1}|^2}2^{\frac{b_{K+1}}{a+\delta}}-\frac{1}{|h_1|^2}-1.\label{neq2}
\end{align}
According to the Jessen inequality
\begin{align}
2^{\frac{b_i}{a+\delta}}&=2^{\frac{a}{a+\delta}\cdot\frac{b_i}{a}+\frac{\delta}{a+\delta}\cdot0}\notag\\
&\leq \frac{a}{a+\delta}2^{\frac{b_i}{a}}+\frac{\delta}{a+\delta}2^0\notag\\
&=\frac{a}{a+\delta}2^{\frac{b_i}{a}}+\frac{\delta}{a+\delta}.\label{neq3}
\end{align}
By substituting (\ref{neq3}) into (\ref{neq2}), we can derive that
\begin{align}
g\left(\mathcal{T}_{d,o}\right)&\leq\sum_{i=1}^K\left(\frac{1}{|h_i|^2}-\frac{1}{|h_{i+1}|^2}\right)\left(\frac{a}{a+\delta}2^{\frac{b_i}{a}}+\frac{\delta}{a+\delta}\right)+\frac{1}{|h_{K+1}|^2}2^{\frac{b_{K+1}}{a+\delta}}-\frac{1}{|h_1|^2}-1\notag\\
&=\frac{a}{a+\delta}\left(\sum_{i=1}^{K-1}\left(\frac{1}{|h_i|^2}-\frac{1}{|h_{i+1}|^2}\right)2^{\frac{b_i}{a}}+\frac{1}{|h_K|^2}2^{\frac{b_K}{a}}-\frac{1}{|h_{K+1}|^2}2^{\frac{b_K}{a}}\right)+\frac{\delta}{a+\delta}\left(\frac{1}{|h_1|^2}-\frac{1}{|h_{K+1}|^2}\right)\notag\\
&\qquad +\frac{1}{|h_{K+1}|^2}2^{\frac{b_{K+1}}{a+\delta}}-\frac{1}{|h_1|^2}-1\notag\\
&\leq \frac{a}{a+\delta}\left(\frac{1}{|h_1|^2}+1-\frac{1}{|h_{K+1}|^2}2^{\frac{b_K}{a}}\right)+\frac{\delta}{a+\delta}\left(\frac{1}{|h_1|^2}-\frac{1}{|h_{K+1}|^2}\right)
+\frac{1}{|h_{K+1}|^2}2^{\frac{b_{K+1}}{a+\delta}}-\frac{1}{|h_1|^2}-1\notag\\
&=-\frac{\delta}{a+\delta}\cdot\left(1+\frac{1}{|h_{K+1}|^2}\right)-\frac{a}{a+\delta}\cdot\frac{1}{|h_{K+1}|^2}2^{\frac{b_K}{a}}+\frac{1}{|h_{K+1}|^2}2^{\frac{b_{K+1}}{a+\delta}}.\label{neq4}
\end{align}

Considering $b_{K+1}=\frac{(1-q)\left(1-(1-q)^{K}\right)}{q}+(1-q)^{K+1}
=b_{K}+(1-q)^{K+1}$, we have
\begin{align}
&~~\frac{1}{|h_{K+1}|^2}2^{\frac{b_{K+1}}{a+\delta}}\notag\\
&=\frac{1}{|h_{K+1}|^2}2^{\frac{b_K}{a+\delta}+\frac{(1-q)^{K+1}}{a+\delta}}\notag\\
&\leq\frac{a}{a+\delta}\frac{1}{|h_{K+1}|^2}2^{\frac{b_K}{a}}+\frac{\delta}{a+\delta}\frac{1}{|h_{K+1}|^2}\cdot2^{\frac{(1-q)^{K+1}}{\delta}}\notag\\
&=\frac{a}{a+\delta}\cdot\frac{1}{|h_{K+1}|^2}2^{\frac{b_K}{a}}+\frac{\delta}{a+\delta}\cdot\left(1+\frac{1}{|h_{K+1}|^2}\right).\label{neq5}
\end{align}
Combining the inequalities in both (\ref{neq4}) and (\ref{neq5}), we may prove that $g(\mathcal{T}_{d,o})\leq 0$. Namely, $\mathcal{T}_{d,p}\leq \mathcal{T}_{d,o}$ holds.

\section{The proof of theorem \ref{the:3}}
We have already proved that, for either centralized case or decentralized case, the concurrent delivery scheme always outperforms orthogonal delivery scheme. In this appendix, we further derive the gap between the decentralized scheme and the centralized scheme when the orthogonal delivery scheme is assumed.

Firstly, let
\begin{align}
\log_2(1+|h_i|^2)&=m+\delta_i;\quad \text{where} \quad 1\leq i\leq K-t,\quad 0=\delta_1\leq \delta_2\ldots\leq \delta_{K-t}\\
b_j&=\left(\frac{K-t}{K}\right)^j; \text{where}\quad 1\leq j\leq K\\
a_i&=\frac{\binom{K-i}{t}}{\binom{K}{t}}; \text{where}\quad 1\leq i\leq K-t
\end{align}
Then $\mathcal{T}_{d,o}$ and $\mathcal{T}_{c,o}$ can be rewritten as
\begin{align}
\mathcal{T}_{d,o}&\leq \sum \limits_{i=1}^{K-t-1} \frac{\left(\frac{K-t}{K}\right)^i}{\log_2\left(1+|h_i|^2\right)}+\sum \limits_{i=K-t}^K \frac{\left(\frac{K-t}{K}\right)^i}{\log_2\left(1+|h_{K-t}|^2\right)} \notag\\
&
=\frac{b_1}{m}+\frac{b_2}{m+\delta_2}+\cdots+\frac{b_{K-t-1}}{m+\delta_{K-t-1}}+\frac{b_{K-t}+\ldots+b_K}{m+\delta_{K-t}},\\
\mathcal{T}_{c,o}&=\frac{a_1}{m}+\frac{a_2}{m+\delta_2}+\cdots+\frac{a_{K-t-1}}{m+\delta_{K-t-1}}+\frac{a_{K-t}}{m+\delta_{K-t}}.
\end{align}
We define the following two functions
\begin{align}
&f_1=(a_1+a_2+\cdots+a_{K-t})\times\left(\frac{b_1}{m}+\frac{b_2}{m+\delta_2}+\cdots+\frac{b_{K-t-1}}{m+\delta_{K-t-1}}+\frac{b_{K-t}+\ldots+b_K}{m+\delta_{K-t}}\right)\label{equ:f1},\\
&f_2=(b_1+b_2+\cdots+b_{K-t}+\cdots+b_K)\times\left(\frac{a_1}{m}+\frac{a_2}{m+\delta_2}+\cdots+\frac{a_{K-t-1}}{m+\delta_{K-t-1}}+\frac{a_{K-t}}{m+\delta_{K-t}}\right)\label{equ:f2}.
\end{align}
In particular, when $K-t=1$
\begin{align}
&f_1=a_1\times \frac{b_1+b_2+\cdots +b_K}{m},\\
&f_2=(b_1+b_2+\cdots +b_K)\times \frac{a_1}{m},
\end{align}
which means $f_1=f_2$.

More generally, let $K-t=l \geq 2$, consider the polynomial expansion for $f_1$ and $f_2$ as below
\begin{align}
f_1=&\frac{a_1b_1}{m}+\frac{a_2b_2}{m+\delta_2}+\cdots+\frac{a_{l-1}b_{l-1}}{m+\delta_{l-1}}+\frac{a_{l}\left(b_{l}+\cdots+b_K\right)}{m+\delta_{l}}\notag\\
&+\frac{a_2b_1}{m}+\frac{a_1b_2}{m+\delta_2}+\frac{a_3b_1}{m}+\frac{a_1b_3}{m+\delta_3}+\cdots+
\frac{a_lb_1}{m}+\frac{a_1b_l}{m+\delta_l}\notag\\
&+\frac{a_3b_2}{m+\delta_2}+\frac{a_2b_3}{m+\delta_3}+\frac{a_4b_2}{m+\delta_2}+\frac{a_2b_4}{m+\delta_4}+\cdots+
\frac{a_lb_2}{m+\delta_2}+\frac{a_2b_l}{m+\delta_l}\notag\\
&\vdots\notag\\
&+\frac{a_lb_{l-1}}{m+\delta_{l-1}}+\frac{a_{l-1}b_l}{m+\delta_l}\notag\\
&+\frac{a_1\left(b_{l+1}+\cdots+b_K\right)}{m+\delta_{l}}+\frac{a_2\left(b_{l+1}+\cdots+b_K\right)}{m+\delta_l}
+\cdots+\frac{a_{l-1}\left(b_{l+1}+\cdots+b_K\right)}{m+\delta_{l}}.
\end{align}
\begin{align}
f_2=&\frac{a_1b_1}{m}+\frac{a_2b_2}{m+\delta_2}+\cdots+\frac{a_{l-1}b_{l-1}}{m+\delta_{l-1}}+\frac{a_{l}\left(b_{l}+\cdots+b_K\right)}{m+\delta_{l}}\notag\\
&+\frac{b_2a_1}{m}+\frac{b_1a_2}{m+\delta_2}+\frac{b_3a_1}{m}+\frac{b_1a_3}{m+\delta_3}+\cdots+
\frac{b_la_1}{m}+\frac{b_1a_l}{m+\delta_l}\notag\\
&+\frac{b_3a_2}{m+\delta_2}+\frac{b_2a_3}{m+\delta_3}+\frac{b_4a_2}{m+\delta_2}+\frac{b_2a_4}{m+\delta_4}+\cdots+
\frac{b_la_2}{m+\delta_2}+\frac{b_2a_l}{m+\delta_l}\notag\\
&\vdots\notag\\
&+\frac{b_la_{l-1}}{m+\delta_{l-1}}+\frac{b_{l-1}a_l}{m+\delta_l}\notag\\
&+\frac{a_1\left(b_{l+1}+\cdots+b_K\right)}{m}+\frac{a_2\left(b_{l+1}+\cdots+b_K\right)}{m+\delta_2}
+\cdots+\frac{a_{l-1}\left(b_{l+1}+\cdots+b_K\right)}{m+\delta_{l-1}}.
\end{align}
\begin{align}
&\Psi=f_1-f_2=\sum  \limits_{i=1}^{l-1} \sum \limits_{j>i}^l \alpha_{ij} +\lambda
\end{align}
Where
\begin{align}
\alpha_{ij}&=\frac{a_jb_i}{m+\delta_i}+\frac{a_ib_j}{m+\delta_j}-\left(\frac{b_ja_i}{m+\delta_i}+\frac{b_ia_j}{m+\delta_j}\right)\label{equ:alphai}, \quad 1\leq i\leq l-1, l\geq j>i, \delta_1=0.\\
\lambda&=\left(\frac{a_1\left(b_{l+1}+\cdots+b_K\right)}{m+\delta_{l}}+\frac{a_2\left(b_{l+1}+\cdots+b_K\right)}{m+\delta_l}
+\cdots+\frac{a_{l-1}\left(b_{l+1}+\cdots+b_K\right)}{m+\delta_{l}}\right) \notag\\
&-\left(\frac{a_1\left(b_{l+1}+\cdots+b_K\right)}{m}+\frac{a_2\left(b_{l+1}+\cdots+b_K\right)}{m+\delta_2}
\cdots+\frac{a_{l-1}\left(b_{l+1}+\cdots+b_K\right)}{m+\delta_{l-1}}\right)
\end{align}
Obviously, $\frac{a_j}{a_i}<\frac{b_j}{b_i}\Rightarrow a_j b_i<a_i b_j$, then let $a_i b_j =a_j b_i +\Delta_{ij}$. We may further derive that
$\alpha_{ij}
\leq 0,   (\delta_j\geq \delta_i).$
Meanwhile, due to $(\delta_2\leq \delta_3 \leq \cdots\leq \delta_l)$, we have $\lambda\leq 0$.
By summing up all the inequalities, we have
\begin{align}
\Psi=\sum  \limits_{i=1}^{l-1} \sum \limits_{j>i}^l \alpha_{ij} +\lambda \leq0
\end{align}
which indicates that
\begin{align}
f_1\leq f_2,
\end{align}
that is to say
\begin{align}
\frac{\mathcal{T}_{d,o}}{\mathcal{T}_{c,o}}\leq \max\left\{\frac{b_1+b_2+\cdots+b_{K-t}+\cdots+b_K}{a_1+a_2+\cdots+a_{K-t}}\right\}.
\end{align}
It is shown in \cite{qifa2016} that
\begin{align}
1\leq \frac{b_1+b_2+\cdots+b_{K-t}+\cdots+b_K}{a_1+a_2+\cdots+a_{K-t}} \leq 1.5,
\end{align}
Hence, the gap between decentralized and centralized orthogonal delivery scheme is
\begin{align}
\frac{\mathcal{T}_{d,o}}{\mathcal{T}_{c,o}}\leq 1.5.
\end{align}

\ifCLASSOPTIONcaptionsoff
  \newpage
\fi

\end{spacing}
\end{document}